\pdfoutput=1
\documentclass[11pt]{article}
\usepackage{jheppub}
\usepackage{epsfig}
\usepackage{amssymb}
\usepackage{amsmath}
\usepackage{tikz}
\usepackage{mathrsfs}
\usepackage{hyperref}
\usepackage{multirow}
\usepackage{scalerel}
\usepackage{mathtools}
\usepackage{textcomp}
\usepackage{color}
\usepackage[all]{xy}
\usepackage{subcaption}

\usetikzlibrary{calc}

\DeclareMathOperator{\Gr}{Gr}
\DeclareMathOperator{\Li}{Li}
\DeclareMathOperator{\sgn}{sgn}

\newcommand{\fwboxL}[2]{\text{\makebox[#1][l]{$#2$}}}

\def\ket#1{\langle #1 \rangle}

\def\x{\mathcal{X}}

\def\xcoords{$\mathcal{X}$-coordinates }
\def\a{\mathcal{A}}

\def\pdfeq#1{\texorpdfstring{$#1$}{a}}

\definecolor{mathematica_magenta}{rgb}{1, 0, 1}

\title{The Two-Loop Remainder Function for \\ Eight and Nine Particles} 

\author{John~Golden$^{1,2}$}
\author{and Andrew~J.~McLeod$^{3}$}

\affiliation{$^1$ Leinweber Center for Theoretical Physics and
Randall Laboratory of Physics, Department of Physics,
University of Michigan
Ann Arbor, MI 48109, USA}

\affiliation{$^2$ Information Sciences (CCS-3), Los Alamos National Laboratory, Los Alamos, NM 87545}

\affiliation{$^3$ Niels Bohr International Academy, Blegdamsvej 17, 2100 Copenhagen, Denmark}

\abstract{Two-loop MHV amplitudes in planar ${\cal N} = 4$ supersymmetric Yang Mills theory are known to exhibit many intriguing forms of cluster-algebraic structure. We leverage this structure to upgrade the symbols of the eight- and nine-particle amplitudes to complete analytic functions. This is done by systematically projecting onto the components of these amplitudes that take different functional forms, and matching each component to an ansatz of multiple polylogarithms with negative cluster-coordinate arguments. The remaining additive constant can be determined analytically by comparing the collinear limit of each amplitude to known lower-multiplicity results. We also observe that the nonclassical part of each of these amplitudes admits a unique decomposition in terms of a specific $A_3$ cluster polylogarithm, and explore the numerical behavior of the remainder function along lines in the positive region.}

\begin{document}
\maketitle

\section{Introduction}

Many of the recent advances in our understanding of scattering amplitudes have come from studying explicit examples. This has been especially true in the planar limit of ${\cal N}=4$ super-Yang-Mills (sYM) theory~\cite{Brink:1976bc,Gliozzi:1976qd}, where nontrivial amplitudes have been computed in six- and seven-particle kinematics through (respectively) seven and four loops~\cite{CaronHuot:2011kk,Dixon:2014iba,Drummond:2014ffa,Dixon:2015iva,Caron-Huot:2016owq,Dixon:2016nkn,Drummond:2018caf,Caron-Huot:2019vjl,Caron-Huot:2020bkp,Dixon:2020cnr}, three-particle form factors have been computed through five loops~\cite{Brandhuber:2012vm,Dixon:2020bbt}, and methods exist for calculating the symbols of several classes of two-loop amplitudes to all particle multiplicity~\cite{CaronHuot:2011ky,Zhang:2019vnm,He:2020vob}. With the benefit of this concrete data, a number of unexpected analytic~\cite{Caron-Huot:2016owq,Caron-Huot:2019bsq}, cluster-algebraic~\cite{Golden:2013xva,Golden:2014pua,Golden:2014xqa,Drummond:2017ssj,Drummond:2019cxm,Arkani-Hamed:2019rds,Henke:2019hve,Mago:2019waa,Gurdogan:2020tip,Mago:2020eua}, number-theoretic~\cite{Caron-Huot:2019bsq}, and positivity~\cite{Arkani-Hamed:2014dca,Dixon:2016apl} properties have been discovered, which have in turn stimulated the development of increasingly advanced computational techniques. The success of these explorations clearly motivates the further study of amplitudes involving eight and more particles, especially as new algebraic and analytic features are known to arise in these higher-point amplitudes. 

A number of perturbative results in planar ${\cal N}=4$ sYM theory are already known for amplitudes involving more than seven particles. In addition to the symbols of the two-loop maximally-helicity-violating (MHV) and next-to-MHV (NMHV) amplitudes, which can be computed at any multiplicity using the methods of~\cite{CaronHuot:2011ky,Zhang:2019vnm,He:2020vob}, special kinematic configurations involving large numbers of particles have also been studied. Such limits include regular polygon configurations~\cite{Brandhuber:2009da,DelDuca:2010zp,Heslop:2010kq,Gaiotto:2010fk}, multi-Regge limits~\cite{Bartels:2008ce,Lipatov:2012gk,Bargheer:2015djt,DelDuca:2016lad,DelDuca:2018raq,DelDuca:2019tur}, and near-collinear limits~\cite{Alday:2010ku,Basso:2013vsa,Basso:2013aha,Basso:2014koa,Basso:2014jfa,Basso:2014nra,Belitsky:2014sla,Belitsky:2014lta,Basso:2014hfa,Belitsky:2015efa,Basso:2015rta,Belitsky:2016vyq}. A handful of Feynman integrals that are expected to contribute to higher-multiplicity amplitudes in this theory have also been computed~\cite{Bourjaily:2018aeq,Henn:2018cdp,He:2020uxy,He:2020lcu,Bourjaily:2021lnz}; however, no complete amplitudes involving more than seven particles are known at function level beyond one loop.

In this paper, we focus on the two-loop MHV amplitudes in planar $\mathcal{N}=4$ sYM theory for eight and nine particles. Once normalized in the infrared by the BDS ansatz~\cite{Bern:2005iz}, these amplitudes respect both a superconformal symmetry and a dual superconformal symmetry~\cite{Drummond:2007au,Drummond:2006rz,Bern:2006ew,Bern:2007ct,Alday:2007he}, the latter of which is naturally associated with light-like polygonal Wilson loops to which these amplitudes are dual~\cite{Drummond:2007au,Alday:2007hr,Drummond:2007aua,Brandhuber:2007yx,Drummond:2007cf,Bern:2008ap,Drummond:2008aq}. The finite part of the $n$-particle amplitude is then encoded in the remainder function $R_n$, which depends only on dual-conformally-invariant (DCI) cross-ratios of Mandelstam invariants and smoothly deforms to the lower-point remainder function $R_{n{-}1}$ in collinear limits.

The symbols of the two-loop eight- and nine-particle remainder functions were computed in~\cite{CaronHuot:2011ky} and are known to exhibit a number of intriguing cluster-algebraic properties. For instance, their logarithmic branch points all coincide with the vanishing loci of cluster coordinates in $\Gr(4,n)$~\cite{Golden:2013xva}, while their Lie cobrackets can be written in terms of Bloch group elements that (termwise) involve only cluster coordinates that appear together in clusters of $\Gr(4,n)$~\cite{Golden:2014pua}. Additionally, the most complicated component of their cobracket---the component encoding the nonclassical part of these functions---can be expressed entirely in terms of simple polylogarithms associated with the $A_2$ or $A_3$ subalgebras of $\Gr(4,n)$~\cite{Golden:2014xqa}. It is further expected that the remaining (classical) contributions to these amplitudes should be expressible in terms of classical polylogarithms with negative cluster coordinate arguments~\cite{Golden:2014xqf}; however, the full functional forms of these amplitudes have for many years remained uncomputed. 

In this paper, we leverage the special cluster-algebraic properties of the two-loop eight- and nine-particle remainder functions to upgrade their symbols to complete analytic functions. We do this using the techniques developed in~\cite{Golden:2014xqa,Golden:2014xqf,Golden:2018gtk,Golden:2019kks}, which involve matching the various components of the remainder function that take different functional forms to appropriate ans\"atze. The terms involving transcendental constants can then be determined from knowledge of the differentials of these remainder functions~\cite{CaronHuot:2011ky,Golden:2013lha} and by taking collinear limits to compare to known lower-multiplicity results.

As part of this analysis, we investigate whether the nonclassical parts of the eight- and nine-particle remainder functions can be decomposed into functions associated with the larger subalgebras of $\Gr(4,8)$ and $\Gr(4,9)$. In~\cite{Golden:2018gtk}, it was shown that the nonclassical part of the seven-particle remainder function could be decomposed into functions associated with the $A_4$, $A_5$, and $D_5$ subalgebras of $\Gr(4,7)$, in addition to the $A_2$ and $A_3$ decompositions found in~\cite{Golden:2014xqa}. We find here that the eight-particle remainder function can be decomposed in terms of the same $A_5$ function that appeared at seven points, and that both the eight- and nine-particle remainder functions admit a unique decomposition in terms of the function $f_{A_3}^{+-}$ that was defined in~\cite{Golden:2014xqa} (which has a different signature under the $A_3$ automorphism group than the original $A_3$ function identified in~\cite{Golden:2014xqa}). Since a unique decomposition of the seven-particle remainder function also exists in terms of $f_{A_3}^{+-}$, we suspect that unique decompositions will exist in terms of this function at all higher $n$.

We go on to make use of our new expressions to explore the behavior of the remainder function in the positive region. We focus on the dihedrally-symmetric point that can be found in this region at each particle multiplicity, and plot the behavior of the remainder function along lines that move away from these symmetric points into successive collinear limits, all the way down to five-point kinematics. This allows us to see the remainder function at each multiplicity smoothly limiting to the appropriate lower-point values along multiple lines, providing a cross-check of our results.
 
The remainder of this paper is organized as follows. We begin in sections~\ref{sec:amplitudes_review} and~\ref{sec:motivic_aspects_polylogs} by briefly reviewing the structure of scattering amplitudes in planar $\mathcal{N}=4$ sYM theory, and the different types of polylogarithmic technology that we will make use of in our analysis. In section~\ref{sec:cluster_algebraic_structure}, we go on to describe the various forms of cluster-algebraic structure enjoyed by two-loop MHV amplitudes in this theory, when considered as functions, symbols, and at the level of their Lie cobracket. Our main results are described in sections~\ref{sec:subalgebra_constructibility} and~\ref{sec:r28_and_r29}. In the former, we describe the different ways in which the nonclassical part of the eight- and nine-particle remainder functions can be decomposed into cluster polylogarithms evaluated on the subalgebras of $\Gr(4,8)$ and $\Gr(4,9)$, and in the latter we describe how the remaining function-level contributions to these amplitudes can be determined. After presenting some numerical results and cross-checks in section~\ref{sec:numerics}, we conclude.

We include explicit polylogarithmic expressions for the eight- and nine-particle remainder functions as ancillary files. As these functions are much too large to be packaged with the arXiv submission, we have made these files available at~\cite{anc_files}.

\section{Amplitudes in Planar $\mathcal{N}=4$ sYM Theory}
\label{sec:amplitudes_review}

We begin by recalling some facts about scattering amplitudes in planar $\mathcal{N} = 4$ sYM theory. In addition to positive- and negative-helicity gluons, this theory involves six scalars and four fermions of each helicity. These fields all transform in the adjoint representation of the $SU(N)$ gauge group, and can be combined into on-shell superfields 
\begin{equation}
\Phi = G^+ + \eta^A \Gamma_A + \frac{1}{2!} \eta^A \eta^B S_{AB} + \frac{1}{3!} \eta^A \eta^B \eta^C \epsilon_{ABCD} \overline{\Gamma}^D + \frac{1}{4!} \eta^A \eta^B \eta^C \eta^D \epsilon_{ABCD} G^-\, ,
\end{equation}
where $G^\pm$ denote positive- and negative-helicity gluons, $\Gamma_A$ and $\smash{\overline{\Gamma}^A}$ identify gluinos and anti-gluinos, and $\smash{S_{AB} = \frac{1}{2} \epsilon_{ABCD} \overline{S}^{CD}}$ represent complex scalars. The R-symmetry indices on these on-shell states transform in the fundamental  representation of $SU(4)$, and are contracted with Gra{\ss}mann variables $\eta^A$ and the Levi-Civita tensor $\epsilon_{ABCD}$. In terms of superfields, amplitudes with different field content can be combined into a single superamplitude $\mathcal{A}_n(\Phi_1,\dots,\Phi_n)$, which naturally stratifies into sectors in which helicity conservation is violated by differing amounts. Namely, 
\begin{align}
\mathcal{A}_n = \mathcal{A}_n^{\text{MHV}} +  \mathcal{A}_n^{\text{NMHV}}  \dots + \mathcal{A}_n^{\overline{\text{MHV}}} \, ,
\end{align}
where the component $\mathcal{A}^{\text{N$^{k}$MHV}}$ collects together all terms of degree $4k{+}8$ in the Gra{\ss}mann variables $\eta^A$. In this paper, we will only study the MHV ($k=0$) component of the superamplitude, which describes (for instance) all-gluon scattering in which all but two gluons have positive helicity (in a convention in which all particles are outgoing). For more background on superfields and superamplitudes, see~\cite{Drummond:2010km,Elvang:2013cua}. 

We further specialize to the planar limit of this theory~\cite{tHooft:1973alw}, where the rank $N$ of the gauge group is sent to infinity while the product $g_{\text{YM}}^2 N$ involving the gauge theory coupling $g_{\text{YM}}$ is held constant. In this limit, multi-trace color structures are suppressed by factors of $1/N^2$, so the leading contribution to the MHV amplitude becomes
\begin{align}
\mathcal{A}_n^{\text{MHV}}  \propto \sum_{\sigma \in S_n/Z_n} \text{Tr}\big( T^{a_{\sigma(1)}} \cdots T^{a_{\sigma(n)}}  \big)  \mathcal{A}_{n,\sigma}^{\text{MHV}} \,+\, \mathcal{O}(1/N^2)\, , 
\end{align}
where $T^{a_i}$ is an $SU(N)$ generator representing the color of the $i^\text{th}$ external particle, and the sum is over all non-cyclically-related permutations of the $n$ external particles. The color-stripped amplitude $\mathcal{A}_{n,\sigma}^{\text{MHV}}$ gathers together all single-trace contributions to the amplitude that are planar with respect to the ordering $\sigma$. Without loss of generality, we focus on the color-stripped amplitude associated with the identity permutation $\sigma = \text{id}$, which collects all contributions in which the external particles are ordered from $1$ to $n$, and drop this label henceforth. 

We study $\mathcal{A}_n^{\text{MHV}}$ at the origin of moduli space, where all fields are massless and the amplitude is infrared-divergent. These divergences are captured by the BDS ansatz $\mathcal{A}_n^{\text{BDS}}$~\cite{Bern:2005iz}, allowing us to define a finite $n$-particle remainder function $R_n$ via the equation
\begin{align}
\mathcal{A}_{n}^{\text{MHV}} = \mathcal{A}_n^{\text{BDS}} \times \exp\left( R_n \right)  \, . \label{eq:remainder_def}
\end{align}
By construction, the BDS ansatz captures the complete one-loop amplitude, so perturbative contributions to the remainder function first occur at two loops:
\begin{equation}
R_n = \sum_{L = 2}^\infty g^{2L} R_n^{(L)} \, ,
\end{equation}
where $\smash{g^2 = \frac{g^2_{\text{YM}} N}{16 \pi^2}}$. In fact, the remainder function is exactly zero for four and five particles due to dual conformal symmetry, while for six or more particles it takes the form of a finite and DCI transcendental function~\cite{Drummond:2006rz, Bern:2006ew, Drummond:2007aua, Bern:2007ct,Nguyen:2007ya, Alday:2007hr, Bern:2008ap, Drummond:2008vq}.

In combination with planarity, dual conformal invariance strongly constrains the kinematic dependence of the nonzero contributions to the remainder function. Namely, these contributions can only depend on DCI cross-ratios of planar Mandelstam invariants $s_{i,\dots,j} = (p_i + p_{i+1} + \dots p_{j})^2$, where $p_i$ is the momentum associated with the $i^{\text{th}}$ external particle. These are often expressed in terms of squared differences of dual coordinates, 
\begin{align}
x_{ij}^2 = (x_j - x_i)^2 = (p_i+ \dots + p_{j-1})^2 = s_{i,\dots,j-1},
\end{align}
where the second equality follows from the definition of the dual coordinates via the relation $p_i = x_{i+1} - x_i$, and all indices are taken modulo $n$ (so $x_{n+i} = x_i$). Ratios of these squared differences are DCI whenever the same indices appear in the numerator and the denominator, for example in $(x_{ij}^2 x_{kl}^2)/(x_{ik}^2 x_{jl}^2)$.  

Four-dimensional dual points $x_i$ naturally live in a six-dimensional embedding space, and there correspondingly exists a Gram determinant relation between any seven of them. A convenient way to satisfy all of these relations is to work in terms of momentum twistors~\cite{Hodges:2009hk}, which provide a minimal parametrization of $n$-particle kinematics. Pragmatically, momentum twistors can be thought of as the columns of a generic $4\! \times \! n$ matrix, modulo the left-action of the projective linear group (encoding dual conformal invariance) and the rescaling of any column by a nonzero complex number (encoding the little group). DCI cross-ratios are related to the entries of this matrix via the map
\begin{align}
x_{ij}^2 = \frac{\text{det}(Z_{i-1} Z_i Z_{j-1} Z_j)}{\text{det}(Z_{i-1} Z_i I_\infty) \text{det}(Z_{j-1} Z_j I_\infty)}  \, ,
\end{align} 
where $Z_i$ is the $i^\text{th}$ column of the momentum twistor matrix $Z$, and $I_\infty$ is the line in momentum-twistor space corresponding to a point at infinity in dual coordinates. While the determinants involving $I_\infty$ break dual conformal invariance, they drop out of DCI cross-ratios. 

More complicated DCI ratios can be formed out of (polynomials of) generic minors of $Z$, which are usually denoted by the four-brackets
\begin{equation} \label{eq:four_bracket_def}
\langle i j k l \rangle = \text{det}(Z_i Z_j Z_k Z_l) \, .
\end{equation}
For instance, the remainder function involves factors usually abbreviated as 
\begin{align}
\langle i (j k)( l m) (n o)\rangle &= 
\langle i j l m \rangle \langle i k n o \rangle - \langle i j n o\rangle \langle i k l m \rangle \, , \label{eq:abbrev_1} \\
\langle i j (k l m) \cap (n o p) \rangle &= 
 \langle i k l m \rangle \langle j n o p \rangle - \langle j k l m \rangle \langle i n o p \rangle \, . \label{eq:abbrev_2}
\end{align} 
Redundancies between (polynomials and ratios of) these minors are captured by Pl\"ucker relations, such as
\begin{equation}
  \label{eq:plucker-rel}
  \ket{abcd} \ket{efcd} = \ket{aecd} \ket{bfcd} - \ket{afcd}\ket{becd} \, .
\end{equation}
All Pl\"ucker relations are satisfied when four-brackets are evaluated on an explicit momentum twistor parametrization. Helpfully, these parameterizations also rationalize large classes of algebraic roots that appear in standard cross-ratio parametrizations~\cite{Bourjaily:2018aeq}. 

Finally, let us mention that the momentum twistor matrix $Z$ can be more formally thought of as an element of the quotient of the Grassmannian $\Gr(4,n)$ by rescalings of its columns~\cite{ArkaniHamed:2012nw},
\begin{equation}
Z \in \Gr(4,n)/GL(1)^{n-1} \, .
\end{equation}
This connection between planar kinematics and $\Gr(4,n)$ will prove important in what follows, as Grassmannians are naturally endowed with a cluster structure~\cite{1021.16017,Golden:2013xva}. In particular, the $\Gr(4,n)$ cluster algebra gives rise to a preferred set of  coordinates (and collections of coordinates) that seem to encode crucial features of the analytic structure of the two-loop remainder function. However, before exploring this connection further, we turn to a description of the types of transcendental functions that appear in these amplitudes.

\section{Motivic Aspects of Multiple Polylogarithms}
\label{sec:motivic_aspects_polylogs}

While the remainder function is expected to depend on increasingly complicated types of functions at large multiplicities and high loop orders~\cite{Paulos:2012nu,CaronHuot:2012ab,Nandan:2013ip,Chicherin:2017bxc,Bourjaily:2017bsb,Bourjaily:2018ycu,Bourjaily:2018yfy,Bourjaily:2019hmc}, it turns out that it can be expressed in terms of just multiple polylogarithms at two loops~\cite{CaronHuot:2011ky}. This is propitious, as our understanding of polylogarithmic functions has advanced greatly over the last two decades. In particular, the analytic structure of these types of functions can be systematically analyzed using the symbol and coaction~\cite{Goncharov:2001iea,Brown:2009qja,Goncharov:2010jf,Brown1102.1312,Brown:2015fyf}, which makes it possible to expose all functional identities between such polylogarithms~\cite{Goncharov:2005sla,2011arXiv1101.4497D,Brown:2011ik,Duhr:2011zq,Duhr:2012fh}.\footnote{Note that the coaction can only be used to find all identities between polylogarithms if the algebraic identities between their symbol letters are also under control, which can prove nontrivial in practice (see for instance~\cite{Bourjaily:2019igt}).} This technology has proven to be increasingly useful as physical constraints on the analytic structure of amplitudes (and related quantities) have become better understood (see for instance~\cite{Bloch:2010gk,Abreu:2014cla,Bloch:2015efx,Abreu:2017ptx,Caron-Huot:2019bsq,Bourjaily:2019exo,Bourjaily:2020wvq,Benincasa:2020aoj}). In the case of the two-loop remainder function, it has also led to the discovery of the various types of cluster-algebraic structure, which we will review in section~\ref{sec:cluster_algebraic_structure}. 

Multiple polylogarithms generalize classical polylogarithms to iterated integrals over more general logarithmic kernels. They often appear in the notation
\begin{equation} \label{eq:G_notation}
G_{a_1,\dots, a_w}(z) = \int_0^z \frac{dt}{t-a_1} G_{a_2,\dots, a_w}(z)\, ,
\end{equation}
where $G(z) = 1$ and the variables $a_i$ and $z$ are complex variables. (In the case of the two-loop remainder function, the $z$ and $a_i$ variables will be rational functions of DCI cross-ratios.) When all of the $a_i$ are zero, this integral diverges and is instead defined to be
\begin{equation}
\quad G_{\fwboxL{27pt}{{\underbrace{0,\dots,0}_{w}}}}(z) = \frac{\log^w z}{w!} \, .
\end{equation}
The number of integrations entering the definition of a polylogarithm is referred to as that function's transcendental weight. 

Multiple polylogarithms can also be defined in terms of nested sums. This gives rise to the alternate notation
\begin{align} \label{eq:Li_notation}
\text{Li}_{n_1,\dots,n_d}(z_1,\dots, z_d) &\equiv \sum_{0 < m_1 < \dots < m_d} \frac{z_1^{m_1} \cdots z_d^{m_d}}{m_1^{n_1} \cdots m_d^{n_d}} \, ,
\end{align}
which reproduces the definition of the classical polylogarithm when the depth $d$ is one. We can relate the sum definition to the integral one using
\begin{align}
\text{Li}_{n_1,\dots,n_d}(z_1,\dots, z_d) = (-1)^d G_{\fwboxL{28pt}{\underbrace{0,\dots,0}_{n_d-1}},\frac{1}{z_d},\fwboxL{13pt}{\,\dots},\fwboxL{29pt}{\,\underbrace{0,\dots,0}_{n_1-1}},\frac{1}{z_1 \cdots z_d}}(1) \, .
\end{align}
We will make use of both of these notations, as they prove useful for different purposes. 

In what follows, we only review the motivic aspects of polylogarithms that we will make use of in our eight- and nine-point computation. In particular, we forego a complete characterization of the coaction, and focus on just the symbol (which corresponds to the maximally iterated coaction) and the Lie cobracket. We refer the interested reader to~\cite{Duhr:2014woa} for a more complete review of these topics.

\subsection{The Symbol}

The symbol map decomposes a generic polylogarithm into a tensor product of logarithms~\cite{Goncharov:2010jf}. It can be iteratively defined in terms of a polylogarithm's total derivatives. For instance, the total differential of the function $G_{a_1, \fwboxL{13pt}{\,\dots}, a_w}(z)$ is given by
\begin{equation} \label{eq:total_differential_G}
d G_{a_1, \fwboxL{13pt}{\,\dots}, a_w}(z) = \sum_{i=1}^{k} G_{a_1,\fwboxL{13pt}{\,\dots},\,\hat{a}_{w-i+1},\fwboxL{13pt}{\,\dots},\,a_w}(z) \ d\log \left( \frac{a_{w-i+1} - a_{w-i}}{a_{w-i+1} - a_{w-i+2}} \right)\, ,
\end{equation} 
where $a_0 \equiv z$ and $a_{w+1} \equiv 0$, and the notation $\hat{a}_{j}$ indicates this index should be omitted~\cite{GoncharovMixedTate,Duhr:2011zq}. The symbol of this function is then defined by promoting the $d\log$s in~\eqref{eq:total_differential_G} to new tensor factors, via the recursive definition  
\begin{equation} \label{eq:symbol_def}
\mathcal{S}\big(G_{a_1, \fwboxL{13pt}{\,\dots}, a_w}(z)\big) \equiv \sum_{i=1}^{w} \mathcal{S}\big(G_{a_1,\fwboxL{13pt}{\,\dots},\,\hat{a}_{w-i+1},\fwboxL{13pt}{\,\dots},\,a_w}(z) \big) \otimes \left( \frac{a_{w-i+1} - a_{w-i}}{a_{w-i+1} - a_{w-i+2}} \right)\, .
\end{equation} 
This recursion terminates once all indices have been dropped, with $\mathcal{S}(G(z)) = 1$.

The algebraic functions that appear in different entries of the symbol are referred to as symbol letters. These letters inherit the distributive properties of (arguments of) logarithms, and can therefore be expanded into a multiplicatively independent basis of symbol letters, referred to as a symbol alphabet. The symbol map thus reduces (potentially complicated) polylogarithmic identities to identities between logarithms, at the cost of losing information about higher-weight transcendental constants and the integration contour of the original polylogarithm. In particular, all terms involving factors of $i\pi$ must be dropped from the symbol, as these terms correspond to deformations of the original integration contour around logarithmic branch points.
 
The symbol captures key information about the analytic structure of polylogarithms, insofar as it encodes their iterated discontinuity structure. Namely, for generic indices $a_i$ and argument $z$, these functions have logarithmic branch points only where the letters in the first entry of their symbol vanish or become infinite. However, the discontinuity of a polylogarithm can itself have new logarithmic branch points when new symbol letters appear in the second entry of its symbol (and similarly for further iterated discontinuities). In this sense, the symbol keeps track of all of a polylogarithm's nonzero sequences of discontinuities. 

\subsection{The Lie Cobracket and Projection Operators}
\label{sec:lie_cobracket}

While the symbol encodes a great deal of information about a function, it does not tell us everything we might want to know; for instance, it does not allow us to compute the numerical value of this function at a given value of its arguments. Thus, we are often in the position of wanting to upgrade the symbol of a function to a complete polylogarithmic expression. While this can prove hard in general, certain information about the polylogarithmic form of a symbol can be discerned with the use of the Lie cobracket~\cite{Golden:2013xva}. 

The cobracket of a symbol $S$ of weight $w$ can be computed as
\begin{equation} \label{eq:cobracket_def}
\delta(S) \equiv \sum_{i=1}^{w-1} (\rho_i \wedge \rho_{w-i})\rho(S) \, ,
\end{equation}
where
\begin{equation}
\rho(s_1 \otimes \cdots \otimes s_w ) = \frac{w-1}{w} \Big(\rho(s_1 \otimes \cdots \otimes s_{w-1}) \otimes s_w - \rho(s_2 \otimes \cdots \otimes s_{w}) \otimes s_1 \Big) \, 
\end{equation}
and $\rho(s_1) \equiv s_1$. The notation in~\eqref{eq:cobracket_def} should be understood to mean that $\rho$ is first applied to $S$, after which each term in the resulting sum is partitioned into a wedge product by splitting up the symbol into its first $i$ and last $w{-}i$ entries; the operator $\rho$ is then applied to the two factors of this wedge product separately. The cobracket thus gives rise to a sum of wedge products involving different transcendental weights. We use the notation $\delta_{i,j}(S)$ to denote all contributions to the cobracket coming from terms that involve a wedge product between factors of weight $i$ and $j$. 

Through weight four, any polylogarithm can be expressed in terms of (products of) classical polylogarithms $\Li_k(z)$ with $k \leq 4$, and nonclassical polylogarithms of the form $\Li_{k,4-k}(x,y)$. The cobracket $\delta$ is useful because it separates out the contribution to a symbol coming from these different types of functions. For instance, the operator $\rho$ that appears in~\eqref{eq:cobracket_def} projects out the part of a symbol that can be written as products of lower-weight polylogarithms. In particular, when it acts on the classical polylogarithm $\Li_k(z)$, it maps it to an element of the Bloch group $\text{B}_k$~\cite{Bloch:2000, Suslin:1990}, or the algebra of polylogarithms modulo identities between classical polylogarithms. We denote these elements by
\begin{align}
 \{ z \}_k  &\equiv \rho(-\text{Li}_k(-z)) \in \text{B}_k, \quad k>1, \\
 \{ z \}_1  &\equiv \rho(\log(z)) \hspace{.675cm} \in \text{B}_1.
\end{align}
At weight four, the cobracket further separates out contributions coming from different types of polylogarithms. More specifically, the $\delta_{2,2}$ component is only sensitive to contributions from the nonclassical polylogarithms $\Li_{k,4-k}(x,y)$, while the $\delta_{3,1}$ component encodes contributions from both these nonclassical polylogarithms and the weight-four classical polylogarithms $\Li_4(x)$~\cite{G91a,2008arXiv0809.3984D,GanglPolylogIdentities,2018arXiv180107816G,2018arXiv180308585G}. 

As we will discuss in more detail in subsequent sections, these facts allow the nonclassical part of the two-loop remainder function to be matched to an appropriate ansatz of polylogarithms at the level of the $\delta_{2,2}$ cobracket. Once a function that reproduces this nonclassical part has been found, we will be interested in matching the different classical contributions to ans\"atze as well. Towards this end, we now describe how these classical contributions can be sequentially isolated, assuming that the nonclassical contribution has been found.

The contribution coming from weight-four classical polylogarithms can be determined by subtracting off the (symbol of the) function that has been chosen to match the nonclassical contribution, and then computing the $\delta_{3,1}$ cobracket component. We note that this classical contribution is not by itself well-defined, since there exist many nonclassical weight-four polylogarithms with the same $\delta_{2,2}$ component but different $\delta_{3,1}$ components. However, any weight-four function whose complete cobracket matches that of the original symbol provides a valid functional representation of the symbol, up to products of lower-weight functions.  

Further projection operators can be defined to isolate and determine the lower-weight products of classical polylogarithms, as described in~\cite{Goncharov:2010jf}. For instance, consider the contribution involving products of weight-two classical polylogarithms, $\Li_2(x) \Li_2(y)$. To determine this component, we define an operator
\begin{equation}
\delta_{\Li_2 \Li_2} (S)= (\rho_2 \otimes \rho_2)\, S \, , \label{eq:proj_op_1}
\end{equation}
which means we apply $\rho$ to the first two entries and last two entries of the symbol separately. It is not hard to check that this operator will project out all weight-four products of classical polylogarithms except for $\Li_2(x) \Li_2(y)$. It will not project out contributions from $\Li_{k,4-k}(x,y)$, but we assume that a function that reproduces these intrinsically weight-four contributions has been subtracted off before computing $\delta_{\Li_2 \Li_2}$. In a similar manner, products of the form $\Li_3(x) \log y$ can be isolated using the operator
\begin{equation}
\delta_{\Li_3 \log} (S)= (\rho_3 \otimes \text{id})\, S \,  \label{eq:proj_op_2} 
\end{equation}
(up to contributions from $\Li_{k,4-k}(x,y)$ or $\Li_4(z)$). 

We can also isolate contributions coming from functions of the form $\Li_2(x) \log y  \log z$. To do so, we define the operator
\begin{equation}
\delta_{\Li_2 \log \log} (S)= (\rho_2 \otimes \text{id})\, S \, . \label{eq:proj_op_3}
\end{equation}
This operator is less precise than the other projection operators; in addition to contributions from $\Li_{k,4-k}(x,y)$ and $\Li_4(z)$, it lets through contributions from $\Li_2(x) \Li_2(y)$ and $\Li_3(x) \log y$. However, we again assume that functional representations of these contributions have been determined and can be subtracted off along with the weight-four contributions. 

Assuming each of the preceding components of the symbol have been matched by appropriate functions, the problem now reduces to matching products of logarithms. This final component is the easiest to upgrade to function level, since it merely requires unshuffling the logarithms whose arguments appear directly in the symbol.  
 
Thus, the projection operators~\eqref{eq:proj_op_1}-\eqref{eq:proj_op_3}, in combination with the different components of the cobracket, allow us to split up the problem of upgrading a symbol to a function into smaller intermediate steps. This can prove necessary when considering large symbols. In extreme cases, the problem of matching a symbol to a function can be split up even further by focusing on terms that involve specific Bloch group elements. We will make use of this strategy in section~\ref{sec:r28_and_r29}, and describe it in more detail there.

\section{Cluster-Algebraic Aspects of \pdfeq{R_n^{(2)}}}
\label{sec:cluster_algebraic_structure}

The analytic structure of the two-loop remainder function has been observed to exhibit many surprising features that seem most elegantly expressed in the language of $\Gr(4,n)$ cluster algebras~\cite{1021.16017}.  In this section, we give a brief overview of some of this structure, with an emphasis on those aspects that will allow us to construct $R_8^{(2)}$\! and $R_9^{(2)}$\! in the next section. We refer the reader to~\cite{Golden:2013xva,Golden:2018gtk} for more comprehensive reviews of cluster algebras and their connection to amplitudes in planar $\mathcal{N} = 4$ sYM theory. 

Let us begin by recalling the definition of the $\Gr(4,n)$ cluster algebra. The initial seed cluster for $\Gr(4,n)$ can be chosen to be~\cite{1088.22009}

\begin{equation}\label{eq:g4n-seed}
\begin{gathered}
\begin{xy} 0;<-.5pt,0pt>:<0pt,-.5pt>::
	 (-100,0) *+{\framebox[8ex]{$\ket{1234}$}} ="-1",
         (300,75) *+{\framebox[5ex]{$f_{21}$}} ="9",
	 (225,75) *+{f_{22}} ="8",
	 (150,75) *+{f_{23}} ="7",
	 (75,75) *+{\color{white} f_{00}} ="6",
	 (0,75) *+{f_{2l}} ="5",
	 (300,0) *+{\framebox[5ex]{$f_{11}$}} ="4",
	 (225,0) *+{f_{12}} ="3",
	 (150,0) *+{f_{13}} ="2",
	 (0,0) *+{f_{1l}} ="0",
	 (75,0) *+{\color{white} f_{00}} ="1",
	 (300,150) *+{\framebox[5ex]{$f_{31}$}} ="14",
	 (225,150) *+{f_{32}} ="13",
	 (150,150) *+{f_{33}} ="12",
	 (75,150) *+{\color{white} f_{00}} ="11",
	 (0,150) *+{f_{3l}} ="10",
	 (0,225) *+{\framebox[5ex]{$f_{4l}$}} ="15",
	 (75,225) *+{\color{white} f_{00}} ="16",
	 (75,0) *+{\cdots} ="01",
	 (75,75) *+{\cdots} ="06",
	 (75,144) *+{\cdots} ="011",
	 (75,225) *+{\cdots} ="016",
	 (150,225) *+{\framebox[5ex]{$f_{43}$}} ="17",
	 (225,225) *+{\framebox[5ex]{$f_{42}$}} ="18",
	 (300,225) *+{\framebox[5ex]{$f_{41}$}} ="19",
	 	"0", {\ar"1"},
		"-1", {\ar"0"},
		"0", {\ar"5"},
		"1", {\ar"2"},
		"6", {\ar"0"},
		"2", {\ar"3"},
		"8", {\ar"2"},
		"2", {\ar"7"},
		"3", {\ar"4"},
		"9", {\ar"3"},
		"3", {\ar"8"},
		"5", {\ar"6"},
		"7", {\ar"8"},
		"6", {\ar"7"},
		"8", {\ar"9"},
		"7", {\ar"12"},
		"13", {\ar"7"},
		"8", {\ar"13"},
		"14", {\ar"8"},
		"5", {\ar"10"},
		"10", {\ar"15"},
		"19", {\ar"13"},
		"12", {\ar"17"},
		"18", {\ar"12"},
		"17", {\ar"11"},
		"13", {\ar"18"},
		"16", {\ar"10"},
		"12", {\ar"13"},
		"13", {\ar"14"},
		"11", {\ar"12"},
		"7", {\ar"1"},
		"10", {\ar"11"},
		"11", {\ar"5"},
		"1", {\ar"6"},
		"12", {\ar"6"},
		"6", {\ar"11"},
		"11", {\ar"16"},
\end{xy}
\end{gathered} 
\end{equation}
where $l=n-4$ and 
\begin{equation}
  f_{i j} =
  \begin{cases}
    \langle i+1, \dotsc, 4, j + 4, \dotsc, i+j+3\rangle, \qquad &i \leq l-j+1,\\
    \langle 1, \dotsc, i+j-l-1, i+1, \dotsc, 4, j+4, \dotsc, n\rangle, \qquad &i >l-j+1.
  \end{cases}
\end{equation}
The four-brackets associated with the nodes of the graph~\eqref{eq:g4n-seed} identify an initial set of cluster $\a$-coordinates, while the arrows define an exchange matrix 
\begin{equation}
b_{i j} = (\# \text{ of arrows}\; i \to j) - (\# \text{ of arrows}\; j \to i) \, .
\label{eq:bijdef}
\end{equation}
The nodes associated with boxed $\a$-coordinates are considered frozen, while all other nodes can be mutated on to generate new clusters. Mutating on node $k$ alters the arrows in a cluster via the following sequence of operations:
\begin{itemize}
	\item[(1)] for each sequence of arrows $i\to k \to j$, add a new arrow $i\to j$,
	\item[(2)] reverse all arrows on edges incident with $k$,
	\item[(3)] remove any two-cycles that have appeared.
\end{itemize}
Algebraically, this generates a new adjacency matrix 
\begin{equation}
  \label{eq:b-mutation}
  b'_{i j} =
  \begin{cases}
    -b_{i j}, &\quad \text{if $k \in \lbrace i, j\rbrace$,}\\
    b_{i j}, &\quad \text{if $b_{i k} b_{k j} \leq 0$,}\\
    b_{i j} + b_{i k} b_{k j}, &\quad \text{if $b_{i k}, b_{k j} > 0$,}\\
    b_{i j} - b_{i k} b_{k j}, &\quad \text{if $b_{i k}, b_{k j} < 0$.}
  \end{cases}
\end{equation}
Mutation also generates new $\a$-coordinates. Denoting the $\a$-coordinate associated with node $i$ by $a_i$, mutation on node $k$ trades the original coordinate $a_k$ for $a_k'$, defined via the relation
\begin{equation}
  \label{eq:a-coord-mutation}
  a_{k} a_{k}' = \prod_{i \vert b_{i k} > 0} a_{i}^{b_{i k}} + \prod_{i \vert b_{i k} < 0} a_{i}^{-b_{i k}},
\end{equation} 
where empty products are set to 1. All other cluster $\a$-coordinates remain unchanged. 

Mutation is an involution, so mutating twice on node $k$ has no net effect. However, mutating on different sequences  of nodes gives rise to new clusters and $\a$-coordinates. The $\Gr(4,n)$ cluster algebra is defined to be the set of all clusters and cluster coordinates that can be generated from~\eqref{eq:g4n-seed} via some sequence of mutations. For $n \le 7$, only a finite number of clusters can be generated, while an infinite number can be generated for $n\ge8$. 

In addition to cluster $\a$-coordinates, cluster algebras give rise to a natural set of cluster $\x$-coordinates. The $\x$-coordinate associated with each mutable node $i$ is given by the ratio of $\a$-coordinates
\begin{equation} \label{eq:x_from_a_coordinates}
	x_i = \prod_j a_j^{b_{ji}}. 	
\end{equation} 
Mutation rules for \xcoords are different than for $\a$-coordinates; when mutating on node $k$, these coordinates change as 
\begin{equation}
  \label{eq:x-coord-mutation}
  x_{i}' =
  \begin{cases}
    x_{k}^{-1}, &\quad i=k,\\
    x_{i} (1+x_{k}^{\sgn b_{i k}})^{b_{i k}}, &\quad i \neq k ,
  \end{cases}
\end{equation}
while the arrows change just as they did for $\a$-coordinates. Note that the translation between $\a$-coordinates and $\x$-coordinates in~\eqref{eq:x_from_a_coordinates} commutes with mutation. 

Because of the inclusion of the frozen nodes in~\eqref{eq:g4n-seed}, the ratios of matrix minors defined by~\eqref{eq:x_from_a_coordinates} are invariant under the rescaling of any column of the underlying matrix. These $\x$-coordinates consequently respect dual conformal symmetry when evaluated on momentum twistor matrices. Since, moreover, the set of $3n{-}15$ $\x$-coordinates found in any cluster of $\Gr(4,n)$ are algebraically independent, cluster algebras give rise to many convenient parameterizations of $n$-particle DCI kinematic space. As we now review, these coordinates also naturally lend themselves to the description of the analytic properties of the two-loop remainder function, making them especially advantageous coordinates in terms of which to formulate these amplitudes. 

\subsection{The Subalgebra Constructibility of \pdfeq{\delta_{2,2}\big(R_n^{(2)} \big)}}

Surprising connections between the analytic structure of $R_n^{(2)}$\! and cluster coordinates on $\Gr(4,n)$ already appear at the level of the cobracket. For instance, in~\cite{Golden:2014pua} it was shown that the cobracket of $R_n^{(2)}$\! can be put in the form
\begin{equation} \label{eq:cobracket_decomp}
\delta\big(R^{(2)}_n\big) = \sum_{i,j} \Big(c_{ij} \, \{x_i\}_2 \wedge \{x_j\}_2 + d_{ij} \, \{x_i\}_3 \wedge \{x_j\}_1 \Big) \, ,
\end{equation}
where $c_{ij}$ and $d_{ij}$ are some rational coefficients, and the arguments $x_i$ and $x_j$ that appear in elements of the Bloch group are cluster $\x$-coordinates. Moreover, it was shown there that the sum over $i$ and $j$ can be restricted to span over only coordinates $x_i$ and $x_j$ that appear together in some cluster of $\Gr(4,n)$.\footnote{Note that this `cobracket-level cluster adjacency' neither implies nor is implied by the similar property of cluster adjacency observed at the level of the symbol in~\cite{Drummond:2017ssj}.} While striking, the physical interpretation of this decomposition remains unclear.

It has also been observed that the $\delta_{2,2}$ component of the cobracket can be decomposed in an entirely different way, into single functions evaluated on different subalgebras of $\Gr(4,n)$~\cite{Golden:2014xqa}. For instance, there always exists a decomposition of the form
\begin{equation} \label{eq:a2_decomp}
	\delta_{2,2} \big(R_n^{(2)}\big) = \!\! \sum_{(x_i\to x_j) \subset \Gr(4,n)} \!\!\! e_{ij} ~\delta_{2,2}\big(f^{--}_{A_2}(x_i \to x_j) \big) \, 
\end{equation}
for some set of rational coefficients $e_{ij}$, where the sum is over a finite number of the $A_2$ subalgebras of $\Gr(4,n)$, each labelled by one of their constituent clusters $x_i \to x_j$. The function $f^{--}_{A_2}$ evaluated on these $A_2$ subalgebras is a weight-four polylogarithm, which (following~\cite{Golden:2018gtk}) we define to be
\begin{align}\label{def:a2-function}
        f^{--}_{A_2}(x_1 \to x_2)  &= \sum_{\text{skew-dihedral}} \bigg[ \Li_{2,2}\left(-\frac{1}{\x_{i-1}},-\frac{1}{\x_{i+1}}\right) - \Li_{1,3}\left(-\frac{1}{\x_{i-1}},-\frac{1}{\x_{i+1}}\right)  \\
        &\hspace{1.1cm} -6 \Li_3\left(-\x_{i-1}\right) \log \left(\x_{i+1}\right) -\frac{1}{2} \log \left(\x_{i-2}\right) \log^2\! \left(\x_i\right) \log \left(\x_{i+1}\right)  {\color{white} \bigg|} \nonumber \\
        &\hspace{1.1cm} +\Li_2(-\x_{i-1}) \Big(3 \log (\x_{i-1})\log (\x_{i+1})+\log \big(\x_{i-2}/\x_{i+2}\big)\log (\x_{i+2})\Big) \bigg], \nonumber
\end{align}
where 
\begin{gather}\label{def:a2-xcoords}
  \x_1 = 1/x_1, \qquad \qquad \x_2 = x_2, \qquad \qquad \x_3 = x_1(1+x_2), \\ 
  \x_4 = \frac{1+x_1+x_1 x_2}{x_2}, \qquad \qquad \x_5 = \frac{1+x_1}{x_1 x_2}, \nonumber
\end{gather}
are five of the cluster $\x$-coordinates generated by the $A_2$ seed cluster $x_1 \to x_2$. The skew-dihedral sum is taken by summing $i$ from 1 to 5, and subtracting off the same quantity in which $\x_i \to \x_{6-i}$.

The function $f^{--}_{A_2}$ has a number of notable features. The arguments of the Bloch group elements that appear in its Lie cobracket and the letters that appear in its symbol are all cluster $\x$-coordinate arguments drawn from $\Gr(4,n)$. It is also a smooth and real-valued function for positive values of the cluster coordinates $x_1, x_2>0$, and it respects the automorphism group of the $A_2$ cluster algebra. This group has two generators,
\begin{align} 
  \sigma_{A_2}:&\quad x_1\to \frac{1}{x_2},~~ x_2\to x_1(1+x_2), \label{eq:a2_automorphism_generator_1} \\
  \tau_{A_2}:&\quad  x_1 \to \frac{x_1 x_2}{1 + x_1}, ~~x_2 \to \frac{1 + x_1 + x_1 x_2}{x_2}, \label{eq:a2_automorphism_generator_2}
\end{align}
both of which map $f_{A_2}^{--}$ back to minus itself (this is the reason for the superscript in the function's name). 

\begin{figure}[t] \centering
  \begin{tikzpicture}
  state/.style={circle, draw, minimum size=3cm}
	\node (H1) at (0cm,.2cm) {\color{white} ${}^i_j F^i_j$};
	\node (H2) at (-2cm,-1.5cm) {\color{white} ${}^i_j F^i_j$};
	\node (H3) at (2cm,-1.5cm) {\color{white} ${}^i_j F^i_j$};
	\node (H4) at (0cm,-3cm) {\color{white} ${}^i_j F^i_j$};
	\node (H5) at (-3cm,-3.4cm) {\color{white} ${}^i_j F^i_j$};
	\node (H6) at (3cm,-3.4cm) {\color{white} ${}^i_j F^i_j$};
	\node (H7) at (0cm,-5.4cm) {\color{white} ${}^i_j F^i_j$};
	\node (P1) at (0cm,.2cm) {$R_7^{(2)}\!\!$};
	\node (P2) at (-1.9cm,-1.5cm) {$f_{D_5}^{---}$};
	\node (P3) at (2cm,-1.5cm) {$f_{A_5}^{--}$};
	\node (P4) at (0cm,-3cm) {$f_{A_4}^{+-}$};
	\node (P5) at (-2.9cm,-3.4cm) {$f_{A_3}^{--}$};
	\node (P6) at (3cm,-3.4cm) {$f_{A_3}^{+-}$};
	\node (P7) at (0cm,-5.4cm) { $f_{A_2}^{--}$};	
	\draw[->] (H1) -- (H3);
	\draw[->] (H1) -- (H2);
	\draw[->] (H2) -- (H4);
	\draw[->] (H2) -- (H5);
	\draw[->] (H3) -- (H4);
	\draw[->] (H3) -- (H6);
	\draw[->] (H5) -- (H7);
	\draw[->] (H4) -- (H7);
	\draw[->] (H6) -- (H7);
\end{tikzpicture}
  \caption{The web of decompositions enjoyed by the nonclassical cluster polylogarithms in terms of which $\delta_{2,2}\big(R^{(2)}_7\big)$\! is subalgebra constructible~\cite{Golden:2018gtk}.}\label{fig:R27_decompositions}
\end{figure}
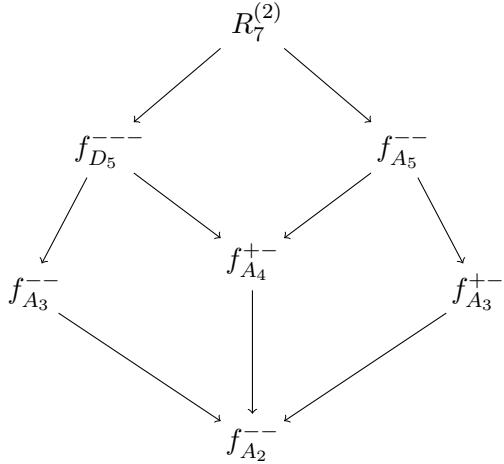

While the decomposition in~\eqref{eq:a2_decomp} is special, it is not unique, nor is this type of decomposition unique to $A_2$ subalgebras. However, only a handful of nonclassical polylogarithms can appear in decompositions of this type, due to the requirement that they respect the automorphism group of the cluster algebra on which they are defined. Despite this, a surprising number of nonclassical decompositions of the two-loop remainder function seem to exist. For instance, the web of nested decompositions depicted in Figure~\ref{fig:R27_decompositions} were found for $\smash{\delta_{2,2}\big( R_7^{(2)}\big)}$ in~\cite{Golden:2018gtk}. In this diagram, each function $f_\mathcal{A}^{s_1 \dots s_n}$ represents a weight-four polylogarithm defined on the cluster algebra $\a$, which has signature $s_i$ under the action of the $i^\text{th}$ generator of the automorphism group of $\a$. (We will give the explicit form of these generators for the $A_n$ cluster algebras in section~\ref{sec:subalgebra_constructibility}; for the $D_5$ case we refer the reader to~\cite{Golden:2018gtk}.) An arrow $\smash{f_\mathcal{A}^{s_1 \dots s_n} \to f_\mathcal{A^\prime}^{s_1^\prime \dots s_m^\prime}}$ indicates that the $\delta_{2,2} \big(\smash{f_\mathcal{A}^{s_1 \dots s_n}}\big)$ can be decomposed into instances of $\smash{f_\mathcal{A^\prime}^{s_1^\prime \dots s_m^\prime}}$ evaluated on the subalgebras $\a^\prime \subset \a$. Note that some sequence of arrows leads from every one of the functions in this diagram to $f_{A_2}^{--}$, implying that they can be all be decomposed in terms of the $A_2$ function defined in~\eqref{def:a2-function}. We refer to the original paper for the definition of the other functions appearing in this diagram.

\subsection{Symbol Alphabets and Cluster Adjacency}

More cluster-algebraic structure can be found in the symbol of the remainder function. In particular, the symbol alphabet of $R_n^{(2)}$\! has been shown to consist of just $\frac{3}{2} n (n - 5)^2$ cluster $\mathcal{A}$-coordinates drawn from $\Gr(4,n)$, implying that the remainder function only develops branch cuts where these cluster coordinates vanish or become infinite~\cite{Golden:2013xva}. It is worth highlighting that, while this observation has also been found to extend to higher loops in the case of $R_6$ and $R_7$~\cite{Caron-Huot:2020bkp,Prlina:2018ukf}, algebraic symbol letters are expected to appear in the remainder function at higher multiplicities. These letters cannot be cluster coordinates, since all cluster coordinates are rational. Even so, it is interesting to note that some of the algebraic letters that appear in planar $\mathcal{N}=4$ sYM theory have been seen to emerge from structures related to Grassmannian cluster algebras~\cite{Drummond:2019cxm,Arkani-Hamed:2019rds,Henke:2019hve,Drummond:2020kqg,Mago:2020kmp,He:2020uhb,Herderschee:2021dez}. 

Further structure can be uncovered by normalizing the amplitude in terms of the BDS-like ansatz~\cite{Alday:2009dv,Yang:2010as}. In this subtraction scheme, it has been observed that symbol letters only appear in adjacent entries when they also appear together in a cluster of $\Gr(4,n)$~\cite{Drummond:2017ssj}. This property, referred to as cluster adjacency, seems to encode the extended Steinmann relations~\cite{Caron-Huot:2018dsv,Caron-Huot:2019bsq}, which generalize the Steinmann relations to all sequential discontinuities~\cite{Steinmann,Steinmann2,Cahill:1973qp}. However, while cluster adjacency is known to imply the extended Steinmann relations~\cite{Golden:2019kks}, the converse has not yet been shown.

Although the decomposition of the amplitude given in equation~\eqref{eq:remainder_def} obscures cluster adjacency, we continue to work with the remainder function for two reasons. First, the BDS-like ansatz does not exist in eight-particle kinematics~\cite{Yang:2010as}. (Although it is possible to define a finite amplitude that respects cluster adjacency and the extended Steinmann relations for this number of particles, it requires giving up dual conformal invariance~\cite{Golden:2018gtk}.) Second, the remainder function is engineered to have smooth collinear limits, which we will see in section~\ref{sec:r28_and_r29} proves to be a more useful property in our analysis than cluster adjacency. 

\subsection{Negative Cluster Coordinate Arguments}
\label{sec:neg_cluster_coodinate_args}

Cluster coordinates also play a special role in the functional form of $R_n^{(2)}$\!. This can be seen explicitly in the cases of $\smash{R_6^{(2)}}$\! and $\smash{R_7^{(2)}}$\!, which turn out to be expressible in terms of polylogarithms (in the sum notation) with negative $\x$-coordinates arguments~\cite{Golden:2013xva,Golden:2014xqf}. The same class of functions also suffices for expressing the nonclassical contribution to $R_n^{(2)}$\! at all multiplicity by virtue of the decomposition in~\eqref{eq:a2_decomp}.\footnote{While we have expressed the weight-one functions in $f^{--}_{A_2}$ as logarithms with positive cluster coordinate arguments, it is not hard to check that these functions can be rewritten in terms of classical polylogarithms with negative cluster coordinate arguments.} It is therefore natural to extend this expectation to the classical component of these functions, and to look for functional representations of $R_n^{(2)}$\! that involve only $f_{A_2}^{--}$ and classical polylogarithms with negative $\x$-coordinates arguments~\cite{Golden:2014xqf}.

The choice of this class of functions has the added benefit that it makes certain properties of the amplitude manifest. In particular, the amplitude is expected to be real-valued in the positive region, which is defined by the inequalities $\langle i j k l \rangle > 0$ for all cyclically ordered $i$, $j$, $k$, and $l$. It is easy to see that the $\a$-coordinates in the initial seed~\eqref{eq:g4n-seed} will all be positive in this region, and that this positivity will be inherited by all further $\a$-coordinates generated by the mutation rule~\eqref{eq:a-coord-mutation}. It follows that all $\x$-coordinates are also positive in this region. As a result, the sum of $A_2$ functions appearing in decompositions of the form~\eqref{eq:a2_decomp} will be manifestly real-valued in the positive region, as $f^{--}_{A_2}(x_1 \to x_2)$ is itself manifestly smooth and real-valued for positive values of $x_1$ and $x_2$. A similar observation holds for the classical polylogarithms $\Li_k(-x)$, which are manifestly real-valued for positive values of $x$.
 
It might seem like restricting our attention to this class of functions will not help in practice, since there are an infinite number of cluster $\x$-coordinates when $n>7$. However, we can restrict our attention to the cluster coordinates that actually appear in the symbol of $R_n^{(2)}$\!. To construct these $\x$-coordinates (since the symbol of $R_n^{(2)}$\! is known only in terms of $\a$-coordinates~\cite{CaronHuot:2011ky}), we begin by listing all DCI cross ratios that can be built out of the $\a$-coordinates that appear in this symbol. We then use the empirical test proposed in~\cite{Golden:2013xva} to determine which of these cross ratios is a genuine $\x$-coordinate. Namely, we select the cross ratios $v$ that have the property that $1+v$ factorizes into a product of $\a$-coordinates, and numerically evaluate $v$, $-1-v$, and $-1-1/v$ at a random point in the positive region. In all known examples, only one of these values will be positive, and will correspond to an $\x$-coordinate. While there is no guarantee that this list of $\x$-coordinates will provide a sufficient set of polylogarithmic arguments for expressing $R_n^{(2)}$\! at function level, we will see in section~\ref{sec:r28_and_r29} that no further $\x$-coordinates are needed in the cases of $R_8^{(2)}$\! or $R_9^{(2)}$\!.

\section{The Subalgebra Constructibility of \pdfeq{R_8^{(2)}} and \pdfeq{R_9^{(2)}}}
\label{sec:subalgebra_constructibility}

Motivated by the web of decompositions found in seven-particle kinematics~\cite{Golden:2018gtk}, we begin our study of $R_8^{(2)}$\! and $R_9^{(2)}$\! by exploring their subalgebra constructibility---that is, we investigate the ways in which $\smash{\delta_{2,2}\big(R_8^{(2)}\big)}$ and $\smash{\delta_{2,2}\big(R_9^{(2)}\big)}$ can be decomposed into polylogarithms evaluated on the subalgebras of $\Gr(4,8)$ and $\Gr(4,9)$. Decompositions of this type were first studied in~\cite{Golden:2014xqa}, where they were found to exist over the $A_2$ and $A_3$ subalgebras of $\Gr(4,n)$ in terms of the functions $f_{A_2}^{--}$ and $f_{A_3}^{--}$. Here, we investigate whether the eight- and nine-particle remainder functions are subalgebra constructible in terms of the other polylogarithms that were found to give rise to nontrivial decompositions of the seven-particle remainder function.

Following~\cite{Golden:2014xqa,Golden:2018gtk}, we refer to the class of polylogarithms that appear in these decompositions of the two-loop remainder function as cluster polylogarithms. More precisely, we adopt the following definition:
\begin{quote}
{\bf Cluster Polylogarithm}: A polylogarithmic function constitutes a cluster polylogarithm on the cluster algebra $\a$ if
\vspace{-.2cm}
 \begin{itemize}
 \item[(i)] its symbol alphabet can be written entirely in terms of $\a$-coordinates of $\mathcal{A}$, 
 \item[(ii)] its Lie cobracket can be expressed in terms of Bloch group elements whose arguments are all $\x$-coordinates of $\mathcal{A}$,
 \item[(iii)] it is invariant under the automorphism group of $\mathcal{A}$, up to a sign.
 \end{itemize}
\end{quote}
The last condition ensures that these functions are well-defined functions on the cluster algebra $\a$, which in particular requires that they return the same value (up to a sign) when evaluated on any of the clusters in $\a$ that share the same directed graph structure.  

A complete classification of the nonclassical cluster polylogarithms that can be defined on the subalgebras of $\Gr(4,n)$ has been carried out through rank five~\cite{Golden:2018gtk} (see also~\cite{Harrington:2015bdt}). In principle, this makes searching for a decomposition of $\smash{\delta_{2,2}\big(R_n^{(2)} \big)}$ over these lower-rank subalgebras easy. One merely needs to evaluate a chosen set of cluster polylogarithms on the appropriate subalgebras of $\Gr(4,n)$, and see if any linear combination of their $\delta_{2,2}$ cobracket components matches $\smash{\delta_{2,2}\big(R_n^{(2)} \big)}$. This procedure can in principle be carried out on any combination of these cluster polylogarithms, but we here restrict our attention to decompositions that only involve a single function.

Of course, an exhaustive search for subalgebra decompositions of this type cannot be carried out when $n>7$, due to the infinite nature of the underlying cluster algebras. We circumvent this problem by restricting our attention to subalgebras of $\Gr(4,n)$ that only involve cluster $\x$-coordinates that appear in the symbol of $R_n^{(2)}$\!. Using the method described at the end of section \ref{sec:neg_cluster_coodinate_args}, we find there are 3176 and 7812 such $\x$-coordinates at eight and nine points, respectively. To generate all subalgebras that can be constructed out of these coordinates, we use the Sklyanin bracket~\cite{Sklyanin:1982tf,GSV}, which allows us compute the number of directed edges between any two cluster $\x$-coordinates when they appear in a cluster together (or alternately tells us that they don't appear together in a cluster). For instance, to search for rank-four subalgebras, we first use the Sklyanin bracket to find all sets of four $\x$-coordinates that can appear together in a cluster of $\Gr(4,n)$; since the Sklyanin bracket also tells us how the nodes associated with these coordinates are connected, we can then mutate on these nodes to determine what type of subalgebra each set generates, and whether this subalgebra involves cluster coordinates beyond those appearing in the symbol of $R_n^{(2)}$\!. For more details on how to compute the Sklyanin bracket between pairs of cluster coordinates, see~\cite{Vergu:2015svm,Golden:2019kks}. 

\begin{table}
\begin{center}
\begin{tabular}{ c |  c | c | c | c | c | c | c }      
 \ & $E_6$ & $D_5$ & $A_5$ & $D_4$ & $A_4$ & $A_3$ & $A_2$  \\
\hline
$\Gr(4,8)$ & 0 & 0 & 56 & 24 & 496 & 1600 & 2240 \\ \hline
$\Gr(4,9)$ & 0  & 0 & 135 & 45 & 1197 & 3936 & 5580 
\end{tabular}
\end{center} 
\caption{The number of subalgebras of $\Gr(4,8)$ and $\Gr(4,9)$ of a given type that can be constructed out of the cluster $\x$-coordinates that appear in the symbols of $R_8^{(2)}$\! and $R_9^{(2)}$\!.}
\label{table:subalgebra_counts}
\end{table}

Using this procedure, we have generated all subalgebras of $\Gr(4,8)$ and $\Gr(4,9)$ through rank five, as well as all $E_6$ subalgebras, that can be constructed solely out of the $\x$-coordinates that appear in the symbols of $R_8^{(2)}$\! and $R_9^{(2)}$\!. We report the number of subalgebras of each type in Table~\ref{table:subalgebra_counts}. The most striking feature of this table is the nonexistence of any $E_6$ or $D_5$ subalgebras that are contructible in this way. This immediately implies that there are no natural decompositions of $\smash{\delta_{2,2}\big(R_8^{(2)}\big)}$ or $\smash{\delta_{2,2}\big(R_9^{(2)}\big)}$ in terms of the function $f_{D_5}^{---}$\!, which gave rise to a unique decomposition of the seven-particle remainder function, or in terms of $R_7^{(2)}$\! itself. However, decompositions into the other functions in terms of which $\smash{\delta_{2,2}\big(R_7^{(2)}\big)}$ was constructible remain possible.

Referring back to Figure~\ref{fig:R27_decompositions}, we see that there are five such functions of interest, associated with the $A_2$, $A_3$, $A_4$, and $A_5$ cluster algebras. As mentioned above, decompositions in terms of two of these functions ($f_{A_2}^{--}$ and $f_{A_3}^{--}$) are already known to exist~\cite{Golden:2014xqa}. However, these decompositions are not unique, whereas the nested decompositions that were found in terms of the remaining three functions ($f_{A_3}^{+-}$, $f_{A_4}^{+-}$, and $f_{A_5}^{--}$) proved to be unique for seven particles. Correspondingly, we would like to answer the question of whether the remainder function is subalgebra constructible in terms of $f_{A_3}^{+-}$, $f_{A_4}^{+-}$, or $f_{A_5}^{--}$ at higher multiplicity---and if so, whether these decompositions remain unique. 
 
\paragraph{The Eight-Particle Remainder Function}~\\[-10pt]

\noindent We first explore this question in eight-particle kinematics. To do so, we construct an ansatz out of each of the functions $f_{A_3}^{+-}$, $f_{A_4}^{+-}$, and $f_{A_5}^{--}$ by evaluating them on the subalgebras of $\Gr(4,8)$ that were found using the methods described above. We then attempt to find $\smash{\delta_{2,2}\big( R_8^{(2)}\big)}$ in the span of the $\delta_{2,2}$ cobracket component of these functions. For instance, to look for a decomposition in terms of the function $f_{A_3}^{-+}$, we see if there exists a rational set of coefficients $c_{ijk}$ such that
\begin{equation}
\delta_{2,2}\big( R_8^{(2)} \big) = \!\! \sum_{(x_i \to x_j \to x_k) \subset \Gr(4,8)} \!\!\! c_{ijk} ~\delta_{2,2}\big( f_{A_3}^{--} (x_i \to x_j \to x_k)\big) \, ,
\end{equation}
where the sum is over all of the $A_3$ subalgebras of  $\Gr(4,8)$ cataloged in Table~\ref{table:subalgebra_counts}, each labelled by one of their constituent clusters $x_i \to x_j \to x_k$. Proceeding in this way, we find novel decompositions of the eight-particle remainder function in terms of both $\smash{f_{A_3}^{+-}}$ and $\smash{f_{A_5}^{--}}$, but no decomposition in terms of $\smash{f_{A_4}^{+-}}$.

To describe these new $A_3$ and $A_5$ decompositions, let us first recall some facts about the automorphism group of the $A_n$ cluster algebra. This group is given by the dihedral group of order $2(3{+}n)$, whose generators can be chosen to be
\begin{equation} \label{eq:def_An_cycle}
  \sigma_{A_n}:\quad x_{k<n} \to \frac{x_{k+1}(1+x_{1,\ldots,k-1})}{1+x_{1,\ldots,k+1}},~~x_n\to\frac{1+x_{1,\ldots,n-1}}{\prod_{i=1}^n x_i} \, 
\end{equation}
and
\begin{equation} \label{eq:def_An_flip}
  \tau_{A_n}: \quad x_1 \to \frac{1}{x_n},~~x_2 \to \frac{1}{x_{n-1}},~\ldots~, \quad x_n\to\frac{1}{x_1} \, ,
\end{equation}
which have lengths $n+3$ and $2$, respectively.\footnote{We note that these generators differ from those given in~\eqref{eq:a2_automorphism_generator_1} and~\eqref{eq:a2_automorphism_generator_2} when $n$ is $2$; however, either set of generators can be used.} The action of these generators is consistent with the notation we have been using for the $A_n$ cluster polylogarithms encountered thus far; that is, the action of $\sigma_{A_n}$ maps the function $f_{A_n}^{s_1 s_2}$ back to itself with an overall sign $s_1$, while the action of $\tau_{A_n}$ maps this function back to itself with an overall sign $s_2$. We refer the reader to~\cite{Golden:2018gtk} for more details on this topic.

The function $f_{A_3}^{+-}$ can be easily defined in terms of $f_{A_2}^{--}$ using the action of the $A_3$ automorphism generators. It is given by 
\begin{align}
f_{A_3}^{+-}(x_1\to x_2\to x_3) &= \sum_{i=1}^6 \sigma_{A_3}^i\big(f_{A_2}^{--}(x_1\to x_2)\big),
\end{align}
where $\sigma_{A_3}^i$ denotes applying the operator $\sigma_{A_3}$ $i$ times. We recall that in seven-particle kinematics, $f_{A_3}^{+-}$ gave rise to a unique decomposition of $\smash{\delta_{2,2}\big( R_7^{(2)}\big)}$. Moreover, as depicted in Figure~\ref{fig:R27_decompositions}, the subalgebras appearing in this seven-particle decomposition came in the right linear combination to combine into specific instances of the function $f_{A_5}^{--}$, defined by
\begin{equation}
	f_{A_3\subset A_5}^{--}\big( x_1\to x_2\to x_3 \to x_4  \to x_5 \big) = - \frac{1}{8}\sum_{A_5^{--}} f_{A_3}^{+-}\left(x_2\to x_3(1+x_4)\to \frac{x_4 x_5}{1+x_4}\right) \, ,
\end{equation}
where we have used the notation
\begin{equation}
\sum_{A_5^{--}} f  \equiv \sum_{i=0}^7 \sum_{j=0}^1 (-1)^{i+j} \ \sigma_{A_5}^i \circ \tau_{A_5}^j \big(f \big)
\end{equation}
to denote the the totally antisymmetric sum over all dihedral images of the $A_5$ cluster algebra.\footnote{Note that this is just one version of the $A_5$ function $f_{A_5}^{--}$ encountered in seven-particle kinematics; there, a variant that could be decomposed into $f_{A_4}^{+-}$ was also found.}

As it turns out, both of these features generalize to eight-particle kinematics: there exists a unique decomposition of $\smash{\delta_{2,2}\big( R_8^{(2)}\big)}$ in terms of $ f_{A_3}^{+-}$ functions, and these $A_3$ functions come in the right linear combination to form instances of $f_{A_3\subset A_5}^{--}$. In terms of the latter functions, the decomposition is given by
\begin{align}\label{eq:r28A5}
\delta_{2,2} \big(R^{(2)}_8\big) &= \frac{1}{40} \sum_{i = 0}^7 \sum_{j=0}^1 \sum_{k=0}^1 \ \sigma_{\Gr(4,8)}^i \circ \tau_{\Gr(4,8)}^j \circ {\mathcal{P}}_{\Gr(4,8)}^j   \Big(\delta_{2,2} \big(F_{A_5} \big)\Big) \, ,
\end{align}
where 
\begin{align} \label{eq:F_A5}
F_{A_5} &= f_{A_3\subset A_5}^{--}\! \left(\tfrac{\langle 1238 \rangle \langle 2345 \rangle}{\langle 1234 \rangle \langle 2358 \rangle} \! 
    \to \! \tfrac{- \langle 1235 \rangle \langle 4568 \rangle}{\langle 5(18)(23)(46) \rangle} \! 
    \to \!\tfrac{\langle 1568 \rangle \langle 2358 \rangle \langle 3456 \rangle}{\langle 1358 \rangle \langle 2356 \rangle \langle 4568 \rangle} \! 
    \to \!\tfrac{\langle 5(18)(23)(46) \rangle}{- \langle 1258 \rangle \langle 3456 \rangle} \! 
    \to \!\tfrac{\langle 1278 \rangle \langle 1358 \rangle}{\langle 1238 \rangle \langle 1578 \rangle}\right)\nonumber\\
   &\quad - \frac12 f_{A_3\subset A_5}^{--} \! \left(\tfrac{\langle 1238\rangle  \langle 1256\rangle }{\langle1235\rangle  \langle 1268\rangle }\! 
    \to \! \tfrac{\langle 1236 \rangle \langle 2345 \rangle}{\langle 1234 \rangle \langle 2356 \rangle} \! 
    \to \! \tfrac{\langle 1235 \rangle \langle 3456 \rangle}{\langle 1356 \rangle \langle 234 5\rangle} \! 
    \to \! \tfrac{\langle 1567 \rangle \langle 2356 \rangle}{\langle 1256 \rangle \langle 3567 \rangle} \! 
    \to \! \tfrac{\langle 1356 \rangle \langle 4567 \rangle}{\langle 1567 \rangle \langle 3456 \rangle}\right)  , 
\end{align}
and the sum ranges over all the dihedral and parity images of $F_{A_5}$ in $\Gr(4,8)$. Explicitly, the generators of the dihedral group act on generic four-brackets as
\begin{align} \label{eq:Gr_4n_dihedral_generators}
  \sigma_{\Gr(4,n)} :\quad  \langle i j k l \rangle \ \ &\to \ \ \langle i{+}1 \,  j{+}1 \,  k{+}1 \,  l{+}1 \rangle \, , \\
  \tau_{\Gr(4,n)} :\quad  \langle i j k l \rangle  \ \ &\to \ \  \langle n{-}i{+}1 \, n{-}j{+}1\, n{-}k{+}1 \, n{-}l{+}1 \rangle \, ,
\end{align}
while parity acts on the four-brackets that appear in the symbol as  
\vspace{1.276cm}
\begin{align} \label{eq:Gr_4n_parity} \ \end{align}
\vspace{-2.82cm}
\begin{align} 
  \mathcal{P}_{\Gr(4,n)} :  \begin{cases} 
	\ket{i \, i{+}1 \, j \, k} \ \ \to\ \ \ket{\bar{i} \, i{+}2}\ket{i \, i{+}1 \, \bar{j}\cap\bar{k}} \\
	\ket{i \, i{+}1 \, \bar{j}\cap\bar{k}} \ \ \to \ \ \ket{j{-}2 \, \bar{j}} \ket{\bar{j} \, j{+}2} \ket{k{-}2 \, \bar{k}} \ket{\bar{k} \, k{+}2} \ket{\bar{i} \, i{+}2}\ket{i \, i{+}1 \, j \, k} \\
	\ket{i \, (i{-}1 \, i{+}1)(j \, j{+}1)(k \,k{+}1)} \\ 
	\qquad \qquad \to \ \ \ket{\bar{i} \, i{+}2} \ket{i{-}2 \, \bar{i}} \ket{\bar{j} \, j{+}2} \ket{\bar{k} \, k{+}2} \ket{i \, (i{-}1 \, i{+}1)(j \, j{+}1)(k \, k{+}1)} \\
	\ket{i \, (i{-}2 \, i{-}1)(i{+}1 \, i{+}2)(j \, j{+}1)} \\ 
	\qquad \qquad \to \ \ \ket{\overline{i{-}2} \, i} \ket{i{-}2 \, \bar{i}} \ket{\bar{i} \, i{+}2} \ket{i \, \overline{i{+}2}} \ket{\bar{j} \, j{+}2} \ket{j \, j{+}1 \, i{-}1\, i{+}1}\, ,
  \end{cases} \nonumber 
\end{align} 
where we have used the notation introduced in~\eqref{eq:abbrev_1} and~\eqref{eq:abbrev_2}, as well as $\bar{i}=(i{-}1\,i\,i{+}1)$. All four-bracket entries should be understood mod $n$. 

The decomposition~\eqref{eq:r28A5} only involves 24 of the $A_5$ subalgebras appearing in Table~\ref{table:subalgebra_counts}, as the $A_5$ cluster algebra appearing in the first term in~\eqref{eq:F_A5} generates a sixteen-orbit under the action of the dihedral group and parity, while the $A_5$ cluster algebra appearing in the second term only generates an eight-orbit (and thus gets counted twice, explaining the relative factor of two). Although we will not discuss collinear limits in detail until section~\ref{sec:collinear_limits}, it is worth commenting on how this decomposition limits to the $A_5$ decomposition of the seven-particle remainder function found in~\cite{Golden:2018gtk}. In the limit in which two adjacent external momenta become collinear, 15 of the 24 $f_{A_3\subset A_5}^{--}$ functions vanish (at the level of the $\delta_{2,2}$ cobracket), while two have residual dependence on the parametrization of the collinear limit (more specifically, on the parameters $\alpha$ and $\beta$ that will appear in equation~\eqref{eq:collinear_parametrization}). However, the two functions with spurious dependence are equal and opposite, and thus cancel out to leave just the 7 $f_{A_3\subset A_5}^{--}$ functions that were observed at seven points.  

\paragraph{The Nine-Particle Remainder Function}~\\[-10pt]

\noindent Despite the existence of unique, nested $A_2 \subset A_3 \subset A_5 \subset \Gr(4,n)$ decompositions at both seven and eight points, no $A_5$ decomposition of the nine-particle remainder function exists (at least in terms of the subalgebras cataloged in Table~\ref{table:subalgebra_counts}). However, there still exists a unique decomposition in terms of $f_{A_3}^{+-}$ functions. It is given by
\begin{align}\label{eq:r29A3}
\delta_{2,2} \big(R^{(2)}_9\big) &= \frac{1}{80} \sum_{i = 0}^8 \sum_{j=0}^1 \sum_{k=0}^1 \ \sigma_{\Gr(4,9)}^i \circ \tau_{\Gr(4,9)}^j \circ {\mathcal{P}}_{\Gr(4,9)}^j   \Big(\delta_{2,2} \big(F_{A_3} \big)\Big) \, ,
\end{align}
where the sum is over all dihedral and parity images of $F_{A_3}$ in $\Gr(4,9)$, and
\begin{align} 
F_{A_3}  &= f_{A_3}^{+-} \left( \tfrac{\langle 1234 \rangle \langle 1256 \rangle}{\langle 1236 \rangle \langle 1245 \rangle} \! 
    \to \! \tfrac{\langle 1246\rangle \langle 2345 \rangle}{\langle 1234 \rangle \langle 2456 \rangle} \! 
    \to \! \tfrac{\langle 1245\rangle \langle 2567 \rangle \langle 3456 \rangle }{\langle 1256 \rangle \langle 2345 \rangle \langle 4567 \rangle} \right) \nonumber \\
&\quad+ f_{A_3}^{+-} \left( \tfrac{\langle 1235 \rangle \langle 1267 \rangle}{\langle 1237 \rangle  \langle 1256 \rangle} \! 
    \to \! \tfrac{\langle 1257 \rangle \langle 2456 \rangle}{\langle 1245 \rangle \langle 2567 \rangle} \! 
    \to \! \tfrac{\langle 1256 \rangle \langle 2678 \rangle \langle 4567 \rangle}{\langle 1267 \rangle \langle 2456 \rangle \langle 5678 \rangle} \right) \nonumber \\
&\quad- f_{A_3}^{+-} \left( \tfrac{\langle 1245 \rangle \langle 1269 \rangle}{\langle 1249 \rangle \langle 1256 \rangle} \! 
    \to \! \tfrac{\langle 1259 \rangle \langle 1267 \rangle \langle 3456 \rangle}{\langle 1269 \rangle \langle 5(12)(34)(67) \rangle} \! 
    \to \! \tfrac{\langle 1256 \rangle \langle 1345 \rangle \langle 4567 \rangle}{\langle 1245 \rangle \langle 1567 \rangle \langle 3456 \rangle} \right) \nonumber \\
&\quad+ f_{A_3}^{+-} \left( \tfrac{\langle 1245 \rangle \langle 2567 \rangle \langle 3456 \rangle}{\langle 1256 \rangle \langle 2345 \rangle \langle 4567 \rangle} \! 
    \to \! \tfrac{\langle 1267 \rangle  \langle 2456 \rangle}{\langle 1246 \rangle \langle 2567 \rangle} \! 
    \to \! \tfrac{\langle 1234 \rangle  \langle 1256 \rangle}{\langle 1236 \rangle \langle 1245 \rangle} \right) \nonumber \\
&\quad- f_{A_3}^{+-} \left(\tfrac{\langle 1256 \rangle \langle 1345 \rangle \langle 4567 \rangle}{\langle 1245 \rangle \langle 1567 \rangle \langle 3456 \rangle} \! 
    \to \! \tfrac{\langle 1249 \rangle \langle 5(12)(34)(67)\rangle}{\langle 1234 \rangle \langle 1259 \rangle \langle 4567 \rangle} \! 
    \to \! \tfrac{\langle 1245 \rangle \langle 1269 \rangle}{\langle 1249 \rangle \langle 1256 \rangle} \right) \\ 
&\quad+ f_{A_3}^{+-} \left( \tfrac{\langle 1256 \rangle \langle 2678 \rangle \langle 4567 \rangle}{\langle 1267 \rangle \langle 2456 \rangle \langle 5678 \rangle} \! 
    \to \! \tfrac{\langle 1278 \rangle \langle 2567 \rangle}{\langle 1257 \rangle \langle 2678 \rangle} \! 
    \to \! \tfrac{\langle 1235 \rangle \langle 1267 \rangle}{\langle 1237 \rangle \langle 1256 \rangle} \right) \nonumber \\
&\quad+ f_{A_3}^{+-} \left( \tfrac{ - \langle 1249 \rangle \langle 1567 \rangle \langle 3467 \rangle}{\langle 4567 \rangle \langle 1(29)(34)(67) \rangle} \! 
    \to \! \tfrac{\langle 1234 \rangle \langle 1269 \rangle \langle 1467 \rangle}{\langle 1249 \rangle \langle 1267 \rangle \langle 1346 \rangle} \! 
    \to \! \tfrac{\langle 3456 \rangle  \langle 1(29)(34)(67) \rangle}{ - \langle 1269 \rangle \langle 1345 \rangle \langle 3467 \rangle} \right) \nonumber \\
&\quad- f_{A_3}^{+-} \left( \tfrac{ - \langle 1249 \rangle \langle 1678 \rangle \langle 3467 \rangle}{\langle 4678 \rangle \langle 1(29)(34)(67) \rangle} \! 
    \to \! \tfrac{\langle 1234 \rangle \langle 1279 \rangle \langle 1467 \rangle}{\langle 1249 \rangle \langle 1267 \rangle \langle 1347 \rangle} \! 
    \to \! \tfrac{\langle 3457 \rangle \langle 1(29)(34)(67) \rangle}{ - \langle 1279 \rangle \langle 1345 \rangle \langle 3467 \rangle} \right) \nonumber \\
&\quad+ f_{A_3}^{+-} \left( \tfrac{ - \langle 1249 \rangle \langle 1789 \rangle \langle 4578 \rangle}{\langle 4789 \rangle \langle 1(29)(45)(78) \rangle} \! 
    \to \! \tfrac{\langle 1245 \rangle \langle 1289 \rangle \langle 1478 \rangle}{\langle 1249 \rangle \langle 1278 \rangle \langle 1458 \rangle } \! 
    \to \! \tfrac{\langle 3458 \rangle \langle 1(29)(45)(78) \rangle}{ - \langle 1289 \rangle \langle 1345 \rangle \langle 4578 \rangle} \right) \nonumber \\
&\quad- \tfrac{1}{3} f_{A_3}^{+-} \left( \tfrac{ - \langle 1249 \rangle \langle 1678 \rangle \langle 4578 \rangle}{\langle 4678 \rangle \langle 1(29)(45)(78) \rangle} \! 
    \to \! \tfrac{\langle 1245 \rangle \langle 1279 \rangle \langle 1478 \rangle}{\langle 1249 \rangle \langle 1278 \rangle \langle 1457 \rangle} \! 
    \to \! \tfrac{\langle 3457 \rangle \langle 1(29)(45)(78) \rangle}{ - \langle 1279 \rangle \langle 1345 \rangle \langle 4578 \rangle} \right) \, . \nonumber 
\end{align}
All of the $A_3$ cluster algebras appearing in $F_{A_3}$ generate eighteen-orbits under the dihedral group and parity, except for the last term which only generates a six-orbit (hence the factor of one third). In collinear limits, 46 of the 168 $f_{A_3}^{+-}$ functions appearing in~\eqref{eq:r29A3} vanish, while 42 of them have spurious dependence on the parametrization of the collinear limit but cancel pairwise. This leaves 80 $A_3$ functions, which assemble into the 24 $A_5$ functions that appear in~\eqref{eq:r28A5} after some cancellations between these $A_5$ functions are taken into account. 

\section{Complete Polylogarithmic Representations of \pdfeq{R_8^{(2)}} and \pdfeq{R_9^{(2)}}}
\label{sec:r28_and_r29}

We now construct complete polylogarithmic representations of $R_8^{(2)}$\! and $R_9^{(2)}$\!, starting from the nonclassical polylogarithms in~\eqref{eq:r28A5} and~\eqref{eq:r29A3} that reproduce the $\delta_{2,2}$ cobracket values of these amplitudes. This is done by isolating their different classical components with the use of the projection operators described in section~\ref{sec:lie_cobracket}, and separately fitting each of these components to an appropriate ansatz. The contributions proportional to transcendental constants can then be determined using knowledge of the differentials $dR_n^{(2)}$\!~\cite{CaronHuot:2011ky,Golden:2013lha}, and by evaluating each function in the collinear limit and comparing it to lower-point results.

\subsection{Classical Contributions}
\label{sec:classical}

Once a polylogarithm with the same nonclassical component as $R_n^{(2)}$\! is known, the contributions from classical polylogarithms can be determined systematically using the methods described in~\cite{Goncharov:2010jf,Golden:2014xqf}. The first step of this procedure is to isolate the additional contribution coming from classical polylogarithms of weight four. As described in section~\ref{sec:lie_cobracket}, this can be done by computing the $\delta_{3,1}$ cobracket component of the difference between the symbol of $R_n^{(2)}$\! and the polylogarithm chosen to match its $\delta_{2,2}$ component. We then upgrade this residual $\delta_{3,1}$ contribution to a sum of weight-four classical polylogarithms by utilizing the expectation that the remainder function is expressible in terms of polylogarithms with negative $\x$-coordinate arguments. That is, we construct an ansatz of weight-four classical polylogarithms with arguments drawn from the list of negative $\x$-coordinates that appear in the symbol of $R_n^{(2)}$\!, and fix the coefficients of this ansatz by requiring it to have the required $\delta_{3,1}$ cobracket value.

The contributions coming from products of lower-weight classical polylogarithms can be determined using a similar procedure---we isolate each functional component using the projection operators defined in section~\ref{sec:lie_cobracket} (after subtracting off all previously-determined contributions), and fit it to an appropriate ansatz. In formulating these ans\"atze, it behooves us to first reduce the number of lower-weight classical polylogarithms we need to include. In particular, while the sets of weight-three and weight-four classical polylogarithms with negative $\x$-coordinate arguments drawn from the symbols of $R_8^{(2)}$\! and $R_9^{(2)}$\! don't prove to be massively overcomplete, the weight-two and weight-one polylogarithms (unsurprisingly) are. Correspondingly, for use in our ans\"atze, we construct a (close to) minimal spanning set of this collection of weight-two classical polylogarithms (modulo products of logarithms). (We don't use a minimal set, because we still want the resulting set of functions to have nice properties under dihedral and parity transformations.) At weight one, we work directly in terms of logarithms of $\a$-coordinates, and only later recombine these logarithms into functions of $\x$-coordinates.  

In this way, starting from the nonclassical polylogarithms in~\eqref{eq:r28A5} and~\eqref{eq:r29A3}, we iteratively build up polylogarithmic functions that reproduce the full symbols of $R_8^{(2)}$\! and $R_9^{(2)}$\!. This requires solving some extremely large systems of linear equations. For instance, the ansatz for the $\Li_2(x) \log y  \log z$ contribution to the nine-particle remainder function, constructed as described above, spans a space of dimension 26904528.\footnote{More precisely, this corresponds to the number of independent DCI degrees of freedom that survive after applying the projection operator~\eqref{eq:proj_op_3}, which maps the dilogarithms to elements of the Bloch group $B_2$.} However, this linear system can be split into different sectors by separating out terms that depend on different pairs of logarithms; these sectors don't mix under the action of the projection operator~\eqref{eq:proj_op_3} as long as all other functional components involving classical polylogarithms of weight two and higher have already been subtracted off. This breaks the problem into 23436 smaller systems of equations, each involving just 1148 degrees of freedom. In the case of the $\Li_2(x) \Li_2(y)$ component, we also use a more tailored ansatz, since this component is known to be expressible in terms of dilogarithms evaluated on a reduced set of $\x$-coordinates (see equation (14) of~\cite{Golden:2014pua}).

After the full symbol of $R_n^{(2)}$\! has been upgraded to a function, the contributions proportional to lower-weight transcendental constants can easily be determined by comparing to the (known) differential of the remainder function, $dR_n^{(2)}$\!~\cite{CaronHuot:2011ky,Golden:2013lha}. Since only $\zeta_2$ appears in these differentials, no other lower-weight constants are needed. Once these $\zeta_2$ terms are incorporated and all logarithms have been recombined to have $\x$-coordinate arguments, our functions involve 4141805 and 51312279 terms at eight and nine points, respectively.\footnote{It is worth noting that the length of these expressions can be brought down multiple orders of magnitude by going to a fibration bais~\cite{Brown:2009qja}. However, since we don't have an \emph{a priori} guess for a minimal set of functions of the latter type that the remainder function can be expressed in terms of, these overly large representations constitute a necessary intermediate step.}

\subsection{Collinear Limits}
\label{sec:collinear_limits}

The only contributions to $R_8^{(2)}$\! and $R_9^{(2)}$\! that remain to be determined are the weight-four constants. Since the remainder function is smooth in collinear limits, namely
\begin{equation} \label{eq:remainder_smooth_collinear_limits}
R_n \xrightarrow[]{p_i || p_{i+1}} R_{n-1} \, ,
\end{equation} 
these contributions can be determined analytically by computing the difference between the collinear limit of each of these functions and the remainder function involving one fewer leg. Thus, we first fix the constant in $R_8^{(2)}$\! by comparing its collinear limit to the known value of $R_7^{(2)}$\!~\cite{Golden:2014xqf}, after which we can similarly fix the constant in $R_9^{(2)}$\!.

Following~\cite{CaronHuot:2011ky}, we parametrize the $R_n^{(2)} \to R_{n-1}^{(2)}$ limit in terms of momentum twistors by
\begin{equation} \label{eq:collinear_parametrization}
Z_n \to Z_{n-1} - \epsilon (\alpha Z_1 + \beta Z_{n-2}) + \epsilon^2 Z_2 \, ,
\end{equation}
where $\epsilon \to 0$, and $\alpha$ and $\beta$ are finite and positive. This parametrization keeps us inside the positive region, and thus ensures we don't cross any branch cuts. Since the $(n{-}1)$-particle amplitude depends on three fewer kinematic degrees of freedom than the $n$-particle amplitude, all three parameters $\epsilon$, $\alpha$, and $\beta$ must drop out in the strict $\epsilon \to 0$ limit. This requirement provides an important cross-check on the functional form of our answers.

To take the collinear limit of our current functions, we substitute the parametrization of $Z_n$ in~\eqref{eq:collinear_parametrization} into all $\x$-coordinates and send $\epsilon \to 0$. This gives rise to a number of polylogarithms with vanishing or divergent arguments. In the case of classical polylogarithms, these arguments can be dealt with using the fact that $\Li_k(0) = 0$, or using the expansions
\begin{align}
\Li_2(-x/\epsilon) &\xrightarrow[]{\epsilon \to 0} - \frac{1}{2} \big(\log(\epsilon) - \log (x) \big)^2  - \zeta_2 \, ,\\
\Li_3(-x/\epsilon) &\xrightarrow[]{\epsilon \to 0}  \frac{1}{6} \big(\log(\epsilon) - \log (x) \big)^3 + 
\big(\log(\epsilon) - \log (x) \big) \zeta_2   \, ,\\
\Li_4(-x/\epsilon) &\xrightarrow[]{\epsilon \to 0}  -\frac{1}{24} \big(\log(\epsilon) - \log (x) \big)^4 - 
  \frac{1}{2} \big(\log(\epsilon) - \log (x) \big)^2 \zeta_2 - \frac{7}{4} \zeta_4  \, ,
\end{align} 
which are valid for positive $x$ and $\epsilon$. The only linear combinations of nonclassical polylogarithms that appear in our functions are $\Li_{2, 2}(x, y) - \Li_{1, 3}(x, y)$, as seen in the first line of~\eqref{def:a2-function}. This function vanishes when either of the arguments $x$ or $y$ goes to zero, even if the other argument diverges at the same rate. As it turns out, we never encounter instances of $\Li_{2, 2}(x, y) - \Li_{1, 3}(x, y)$ in which one of the arguments diverges while the other remains finite (or itself diverges), so we never need to expand these functions around infinity. 

In order to see the dependence on $\epsilon$, $\alpha$, and $\beta$ drop out in this limit, we must take into account a large number of nontrivial polylogarithmic identities, such as
\begin{align}
\Li_{2, 2}(-x, -1/x) - \Li_{1, 3}(-x, -1/x) &=  \Li_{2, 2}(-1/x, -x) - \Li_{1, 3}(-1/x, -x)  \nonumber \\
&\quad  + 10 \Li_4(-x)  - \frac{1}{2} \Li_2(-x) \left( \log(x)^2 + 14 \zeta_2 \right)   \\ &\quad  - \frac{1}{6} \Li_1(- x) \log(x) \left( \log(x)^2 + 6 \zeta_2 \right) \, .   \nonumber
\end{align}
Rather than find all such identities explicitly, we evaluate all $\x$-coordinates on an explicit momentum twister parameterization, which allows us to express our functions in terms of a fibration basis in these parameters~\cite{Brown:2009qja}. (This can be done using publicly-available codes such as~\cite{Panzer:2014caa,Duhr:2019tlz}.) In addition to seeing the parameters $\epsilon$, $\alpha$, and $\beta$ drop out, this makes the comparison to the lower-point remainder function easier, since we can express both functions in the same fibration basis.

Using this procedure, we have analytically determined the constant contributions to $R_8^{(2)}$\! and $R_9^{(2)}$\!, both of which are proportional to $\zeta_4$ (as expected). The specific coefficients of $\zeta_4$ in these functions can be found in the ancillary files, but aren't meaningful in isolation (as they can be shifted with the use of polylogarithmic identities). This completes our determination of the two-loop eight- and nine-particle MHV amplitudes.

\section{Numerical Results and Cross-Checks}
\label{sec:numerics}

One of the benefits of having a full polylogarithmic representation of the $n$-particle remainder function is that it makes it easier to get numerical results for the corresponding amplitude. Studying such numerics helps us better understand the behavior of these amplitudes, and in the present context also provides an important cross-check on our analytic determination of the $\zeta_4$ contributions to $R_8^{(2)}$\! and $R_9^{(2)}$\!. As we will see below, our new representations of the eight- and nine-particle remainder functions indeed limit to the correct numerical values along lines that approach collinear limits.

We restrict our numerical explorations to the positive region, where our functional representations are manifestly real. In~\cite{Galashin:2017onl}, it was shown that this region is homeomorphic to a ball centered at the point corresponding to the four-bracket values
\begin{align}
\langle i j k l \rangle \big|_{\text{sym}(n)} &= \sin\left(\frac{(l - k) \pi}{n} \right) \sin\left(\frac{(l - j) \pi}{n} \right) \sin\left(\frac{(l - i) \pi}{n}\right) \nonumber \\
&\qquad \times \sin\left(\frac{(k - j) \pi}{n}\right) \sin\left(\frac{(k - i) \pi}{n}\right) \sin\left(\frac{(j - i) \pi}{n}\right) \, . \label{eq:symmetric_point}
\end{align}
This point is dihedrally symmetric in $n$-particle kinematics. Drawing upon known representations for the two-loop six- and seven-particle remainder functions~\cite{DelDuca:2009au,DelDuca:2010zg,Goncharov:2010jf,Golden:2014xqf}, we have evaluated $R_n^{(2)}$ for all $n \leq n^\prime \leq 9$ on the four-bracket values $\langle i j k l \rangle \big|_{\text{sym}(n^\prime)}$ using {\sc GiNaC}~\cite{Bauer:2000cp,Vollinga:2004sn}; the results are presented in Table~\ref{table:symmetric_point_numerics}. While the  remainder function does not in general evaluate to a linear combination of nice transcendental constants at these points (such as multiple polylogarithms evaluated at sixth roots of unity), we can still use these numerical values to check that the $n$-particle remainder function limits to the correct lower-point values when $Z_n \to Z_{n-1}$. 

\begin{table}[!t]
\begin{center}
\renewcommand{\arraystretch}{1.2}
\begin{tabular}{| l || c | c | c | c |}
\hline
\ & $\langle i j k l \rangle \big|_{\text{sym}(9)}$ & $\langle i j k l \rangle \big|_{\text{sym}(8)}$ & $\langle i j k l \rangle \big|_{\text{sym}(7)}$ & $\langle i j k l \rangle \big|_{\text{sym}(6)}$ \\[.12cm]
\hline\hline
$R^{(2)}_6$  & $4.4955718025$ & $4.4596240384$ & $4.4070349650$ & $4.3566635946$  \\
$R^{(2)}_7$  & $10.6344626298$ & $10.5308343238$  & $10.4368451968$ & \ \\
$R^{(2)}_8$  & $17.5132409036$ & $17.3847911942$  & \  & \ \\
$R^{(2)}_9$  & $24.8343799325$ & \  & \  & \ \\
\hline
\end{tabular}
\caption{The $n$-point remainder function evaluated at the totally dihedral point $\langle i j k l \rangle \big|_{\text{sym}(n^\prime)}$ for $n \leq n^\prime \leq 9$.}
\label{table:symmetric_point_numerics}
\end{center}
\end{table}

To check this limiting behavior, we parametrize a line between the totally-dihedral point in $n$-particle kinematics, and the point in $(n{-}1)$-particle kinematics that corresponds to sending $Z_n \to Z_{n-1}$. In order to do so, we first solve
\begin{equation} \label{eq:totally_symmetric_parameters}
\Big[ \langle i j k n \rangle = a\, \langle i j k \, n{-}1 \rangle - b\, \langle i  j  k 1 \rangle - c\, \langle i j k \, n{-}2 \rangle + d\, \langle i j k 2 \rangle \Big]_{\text{sym}(n)} 
\end{equation}
for $a$, $b$, $c$, and $d$ by plugging in various values of $i$, $j$, and $k$ and evaluating these four-brackets on~\eqref{eq:symmetric_point}. This just amounts to expanding the momentum twistor $Z_n$ in terms of the twistors $Z_1$, $Z_2$, $Z_{n-2}$, and $Z_{n-1}$ at this kinematic point, and thus any four independent equations will give the same result. We then adopt the parametrization
\begin{equation} \label{eq:line_parametrization}
Z_n = a\, Z_{n-1} - x (b\, Z_1 + c\, Z_{n-2}) + x^2 \, d \, Z_2 \, , \qquad 0 < x \leq 1
\end{equation}
which (using the values for $a$, $b$, $c$, and $d$ determined by~\eqref{eq:totally_symmetric_parameters}) reproduces the dihedrally symmetric point~\eqref{eq:symmetric_point} when $x$ is 1, and approaches the collinear limit in an appropriate manner as $x \to 0$. In particular, the $n$-particle remainder function should numerically approach the value of the $(n{-}1)$-point remainder function evaluated on $\langle i j k l \rangle \big|_{\text{sym}(n)}$ as $x \to 0$.

More generally, the parametrization in~\eqref{eq:line_parametrization} can be used to move between any initial point in $n$-particle kinematics and the point in $(n{-}1)$-particle kinematics that corresponds to forgetting the twistor $Z_n$. One merely solves the constraint
\begin{equation}
\Big[ \langle i j k n \rangle = a\, \langle i j k \, n{-}1 \rangle - b\, \langle i  j  k 1 \rangle - c\, \langle i j k \, n{-}2 \rangle + d\, \langle i j k 2 \rangle \Big] 
\end{equation}
for $a$, $b$, $c$, and $d$ on the four-bracket values corresponding to the desired initial point in $n$-particle kinematics, instead of at the dihedrally-symmetric point. In this way, starting from any initial kinematic point, we can use the parametrization~\eqref{eq:line_parametrization} to iteratively take collinear limits until we approach a point in five-particle kinematics, where the remainder function must vanish. Note that we always apply the parametrization~\eqref{eq:line_parametrization} to the momentum twistor with the highest index $n$, so this prescription associates a unique line from every kinematic point (in arbitrarily high-point kinematics) down to five- (or lower-)point kinematics.

\begin{figure}[t]
\captionsetup[subfigure]{labelformat=empty}
\begin{center}
  \begin{subfigure}[b]{0.85\textwidth}
    \includegraphics[width=\textwidth]{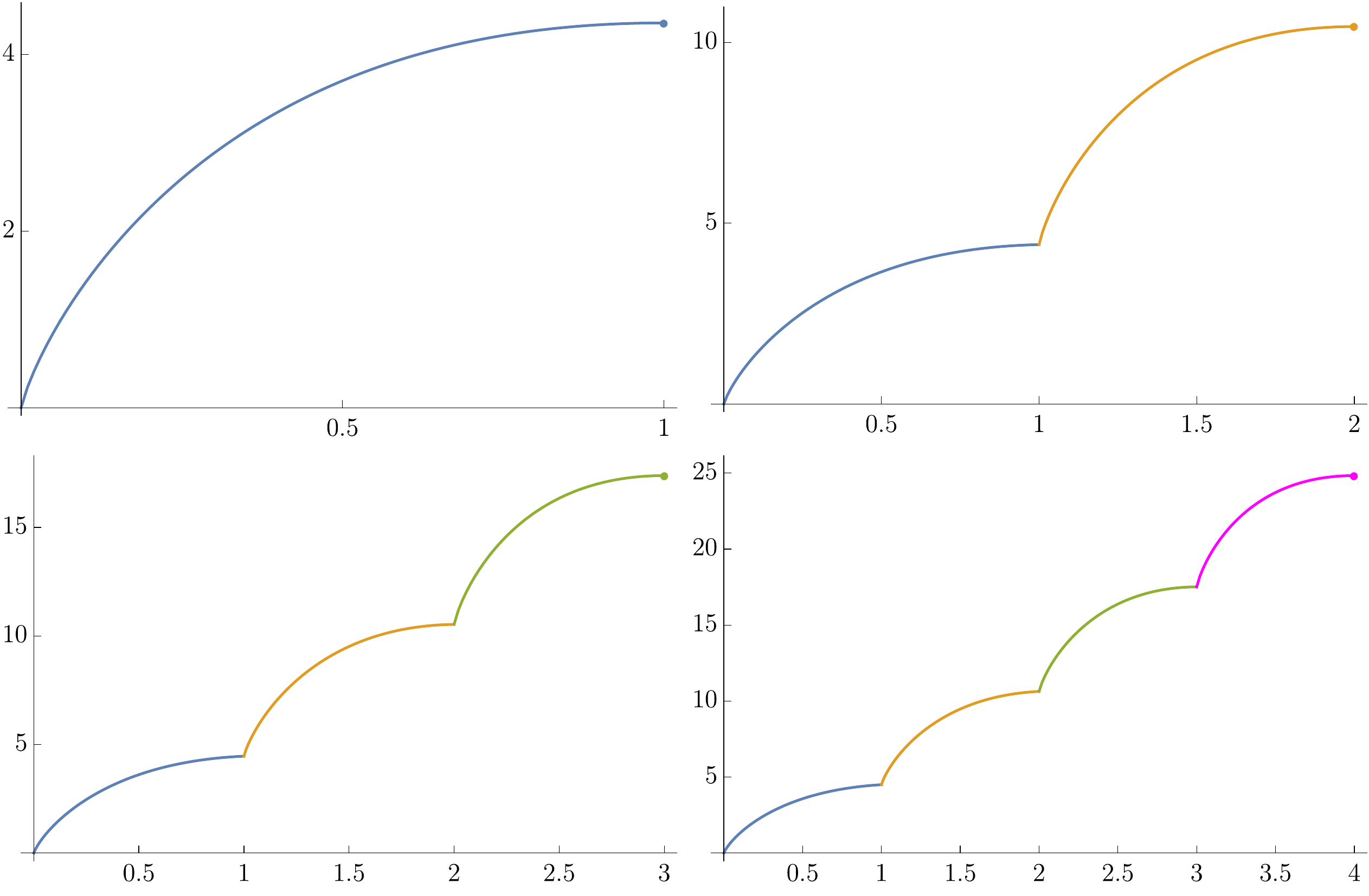}
  \end{subfigure}
\  %
  \begin{subfigure}[t]{0.1\textwidth} \vspace{-7cm}
    \includegraphics[width=\textwidth]{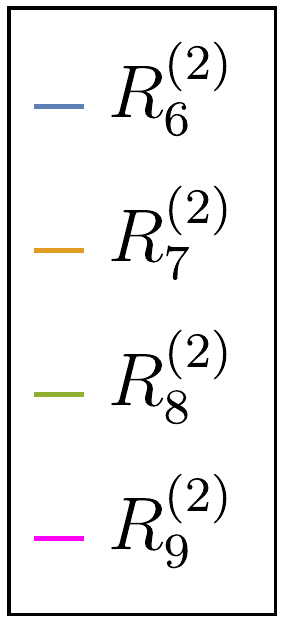}
  \end{subfigure}
    \caption{The remainder function evaluated at the dihedrally-symmetric point in six- through nine-particle kinematics and then iteratively taken into collinear limits using the parametrization given in~\eqref{eq:line_parametrization}.}
    \label{fig:symm_to_collinear_lines}
\end{center}
\end{figure}

\begin{figure}[t]
\captionsetup[subfigure]{labelformat=empty}
\begin{center}
  \begin{subfigure}[b]{0.6\textwidth}
    \includegraphics[width=\textwidth]{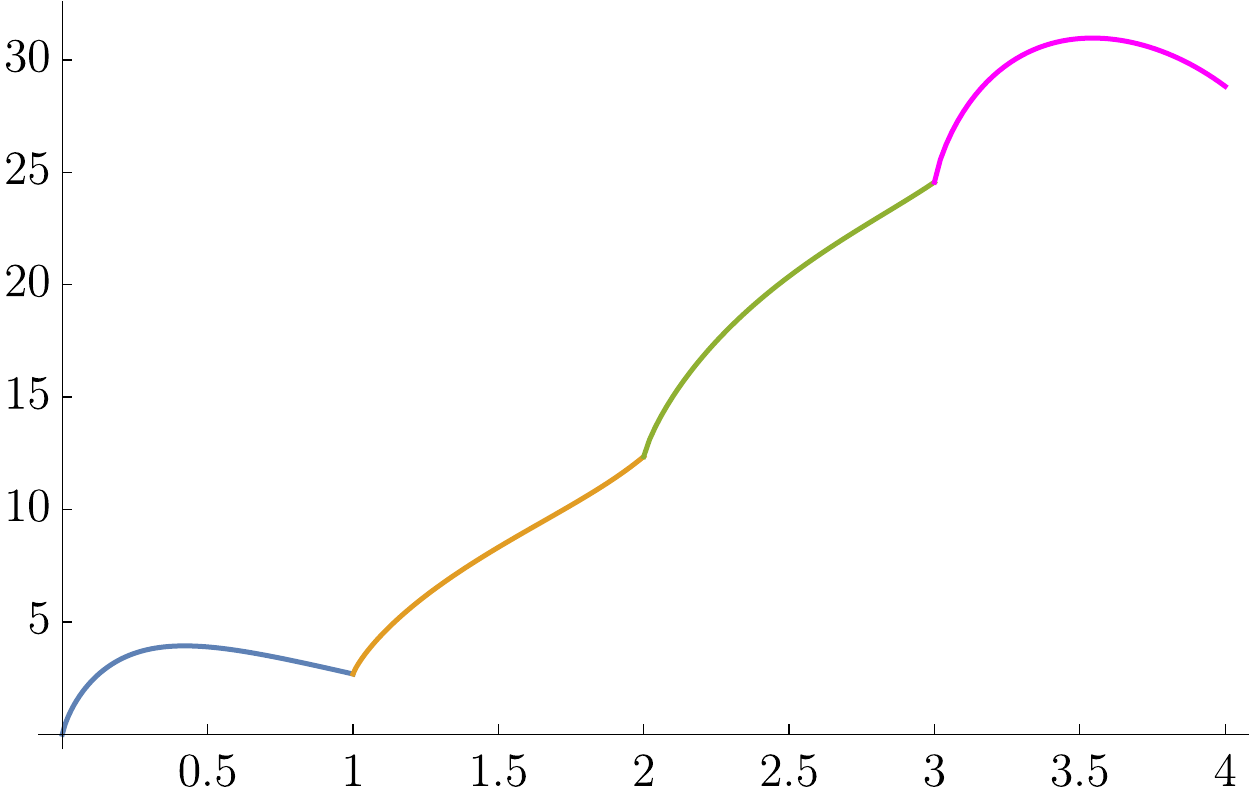}
  \end{subfigure}
\  %
  \begin{subfigure}[t]{0.1\textwidth} \vspace{-5cm}
    \includegraphics[width=\textwidth]{symmetric_lines_label.pdf}
  \end{subfigure}
    \caption{The remainder function evaluated on the line associated with the randomly chosen kinematic point given by the momentum twistor matrix in~\eqref{eq:random_kinematic_point}.}
    \label{fig:random_point_to_collinear_lines}
\end{center}
\end{figure}

We now focus on the lines this prescription associates with the dihedrally-symmetric points in six-, seven-, eight-, and nine-particle kinematics. The value of the remainder function along these lines is plotted in Figure~\ref{fig:symm_to_collinear_lines}. Since these lines are parametrized piecewise (with different segments corresponding to kinematics involving different numbers of particles), a kink naturally appears each time the parametrization~\eqref{eq:line_parametrization} approaches a collinear limit. However, what is important is that the endpoints of the line segments in adjacent kinematic regions limit to a common value, in accordance with~\eqref{eq:remainder_smooth_collinear_limits}; in particular, the values at which these line segments meet are precisely those reported in Table~\ref{table:symmetric_point_numerics}.\footnote{We have shifted the line segments appearing in Figures~\ref{fig:symm_to_collinear_lines} and~\ref{fig:random_point_to_collinear_lines} along the $x$ axis in order to emphasize that their endpoints appropriately limit to common values; for instance, the values of $R_9^{(2)}$\! corresponding to $0 < x \leq 1$ in~\eqref{eq:line_parametrization} are given in the range $3 < x \leq 4$ in these plots.} For instance, the line associated with the dihedrally-symmetric point in nine-particle kinematics is shown in the bottom right plot of Figure~\ref{fig:symm_to_collinear_lines}. In this plot, the rightmost point depicts the value of $R_9^{(2)}$\! at the dihedrally-symmetric nine-particle point, while the kinks one encounters as one moves to the left occur at the values of $R_8^{(2)}$\!, $R_7^{(2)}$\!, and $R_6^{(2)}$\! evaluated on the same four-bracket values, as given in the first column of Table~\ref{table:symmetric_point_numerics}. The kinks in the other three plots, which depict the lines that start at the dihedrally-symmetric points in six-, seven-, and eight-particle kinematics, occur at the values given in the other three columns. As expected, these lines all vanish in the collinear limit that approaches five-point kinematics. The fact that each of the (differently-colored) line segments in these plots limits to the correct value of the lower-point remainder function provides a nontrivial check on our determination of $R_8^{(2)}$\! and $R_9^{(2)}$\!.  

In Figure~\ref{fig:symm_to_collinear_lines}, we see that the remainder function decreases monotonically as one moves away from the dihedrally-symmetric points given in~\eqref{eq:symmetric_point} along the lines parametrized by~\eqref{eq:line_parametrization}.\footnote{We note, however, that the remainder function is not at a local maximum at these dihedrally-symmetric points.} Moreover, the line segments appearing in this figure all take approximately the same shape. This can be contrasted with Figure~\ref{fig:random_point_to_collinear_lines}, where we have plotted the remainder function on the line starting at a randomly chosen point in nine-particle kinematics, given by the momentum twistor matrix
\begin{equation} \label{eq:random_kinematic_point}
\left(
\begin{array}{ccccccccc}
 1 & \frac{57}{16} & \frac{53773}{1024} & \frac{8965489}{94208} & \frac{4259781}{94208} & \frac{123975}{23552} & 0 & 0 & 0 \\
 0 & 1 & \frac{30377}{32} & \frac{43566331}{20608} & \frac{24024151}{20608} & \frac{978025}{5152} & \frac{13775}{644} & 0 & 0 \\
 0 & 0 & 1 & \frac{16043}{7140} & \frac{8903}{7140} & \frac{377}{1785} & \frac{1276}{41055} & \frac{58}{8211} & 0 \\
 0 & 0 & 0 & 1 & 1 & 1 & 1 & 1 & 1 \\
\end{array}
\right).
\end{equation}
We see there that, while the endpoints of adjacent line segments appropriately limit to common values, the remainder function exhibits more erratic behavior in general. However, the value of the remainder function still increases, on average, as one moves to higher $n$. It would be interesting to know if this is a generic feature of these amplitudes, or perhaps an artifact of the parametrization~\eqref{eq:line_parametrization}.

As an additional cross-check, we have evaluated our expressions for $R^{(2)}_8$\! near a singular, codimension-six surface analogous to the collinear-origin surface of the seven-point amplitudes studied in~\cite{Dixon:2020cnr}, which contains an origin similar to the six-point origin studied in~\cite{Basso:2020xts}, and we have found exact (analytic) agreement with the expression independently computed by~\cite{Lance_to_appear}. This check, in combination with the numerical checks presented above, give us high confidence in our results.\footnote{It would also have been interesting to compare to the numerical results of~\cite{Anastasiou:2009kna,Brandhuber:2009da} or the analytic results in special kinematics studied in~\cite{DelDuca:2010zp,Heslop:2010kq,Gaiotto:2010fk}, but these configurations are either off the support of Gram determinant constraints or far outside the positive region, making them harder to access with our results.}

\section{Conclusion}

In this paper, we have finished the computation of the two-loop eight- and nine-particle remainder functions, upgrading their symbols to complete polylogarithmic functions. To do so, we have made use of the extensive and well-studied cluster-algebraic structure enjoyed by these amplitudes, in particular making use of the algorithm described in~\cite{Golden:2014xqf}. Our representations of these amplitudes are manifestly real in the positive region, making them well-suited to numerical evaluation there. Leveraging this fact, we have computed the value of the remainder function at the unique diherally-symmetric points that appear in this region at each multiplicity, and have explored the behavior of the six-, seven-, eight-, and nine-particle amplitudes along lines that move away from these points into (iterated) collinear limits.

We have also explored the subalgebra constructibility of the eight- and nine-particle amplitudes, extending the analyses of~\cite{Golden:2014xqa,Golden:2018gtk}. While exhaustive surveys for such decompositions are no longer possible for these particle multiplicities (since the underlying cluster algebras are infinite), we have searched for all possible decompositions involving just the $\x$-coordinates that appear in the symbol of each function. We have found that a unique $A_5$ decomposition of this type exists for the eight-particle remainder function, similar to what was found at seven points, but that no such $A_5$ decomposition exists at nine points. More encouragingly, we have observed that a unique $A_3$ decomposition in terms of the function $f_{A_3}^{+-}$ exists for seven, eight, and nine particles. We suspect this points to the existence of a unique $f_{A_3}^{+-}$ decomposition at all particle multiplicities; if so, it would be interesting to find a closed-form expression for this decomposition that was valid at all $n$.\footnote{Perhaps the recently-introduced class of `clean single-valued polylogarithms' could be useful for finding such an expression~\cite{Charlton:2021uhu}.}

It should also be possible to use the methods we have employed here (and that were first proposed in~\cite{Golden:2014xqf}) to compute the remainder function at ten and higher points. The main barrier to doing so is a computational one. Even if a closed-form polylogarithmic expression were found that reproduced the nonclassical part of the two-loop remainder function at all $n$, the additional contributions coming from classical polylogarithms would need to be determined. As evidenced by the sizes of the spaces we encountered at eight and nine points (some of which were reported in section~\ref{sec:classical}), this problem will quickly grow unfeasible. However, it may be that there exists further all-multiplicity structure in these classical components that only becomes visible once a more canonical representation of the nonclassical part of the remainder function is utilized. Such structure could then reduce the computational burden of this part of the procedure. We leave this question to future work.

Finally, while many of the features that we have leveraged in our analysis seem to be unique to the two-loop MHV amplitudes of planar $\mathcal{N}=4$ sYM, it remains worth considering whether similar techniques could be applied to closely-related classes of amplitudes. For instance, the symbols of the NMHV amplitudes in this theory can now be computed at any multiplicity~\cite{Zhang:2019vnm,He:2020vob}, while the integrand of the nonplanar two-loop MHV amplitude is also known to all multiplicity, in terms of a basis of just six Feynman integrals~\cite{Bourjaily:2019iqr,Bourjaily:2019gqu}. Although neither of these classes of amplitudes is expected to exhibit the same type of cluster-algebraic structure as the amplitudes studied in this paper, it remains possible that their Lie cobrackets could exhibit a different type of structure that could be leveraged in a similar way.

\section*{Acknowledgements}

We thank Lance Dixon, Mark Spradlin, and Chi Zhang for illuminating discussion and comments on the manuscript. We also thank Lance Dixon for sharing his expression for one of the codimension-six limits of the eight-particle amplitude, to appear in~\cite{Lance_to_appear}. This project has received funding from an ERC Starting Grant \mbox{(No.\ 757978)}, a grant from the Villum Fonden, and a Carlsberg Postdoctoral Fellowship (CF18-0641) (AJM), and the support of a Van Loo Postdoctoral Fellowship (JG).

\bibliographystyle{JHEP}
\bibliography{subalgebras}

\end{document}